\documentclass[%
reprint,
 amsmath,amssymb,
 aps,
]{revtex4-2}

\usepackage{graphicx}% Include figure files
\usepackage{dcolumn}% Align table columns on decimal point
\usepackage{bm}% bold math
\makeatletter
\renewcommand{\fnum@figure}{Figure \thefigure}
\makeatother
\usepackage[version=4]{mhchem}%Chemical formulas

\begin{document}
\preprint{APS/123-QED}

\title{Autonomous Engulfment of Active Colloids by Giant Lipid Vesicles}

\author{Florent Fessler}
\email{florent.fessler@ics-cnrs.unistra.fr, stocco@unistra.fr}
\affiliation{ Institut Charles Sadron, CNRS UPR-22, 23 rue du Loess, Strasbourg, France}
 \author{Martin Wittman}
\affiliation{Physical Chemistry, TU Dresden, Zellescher Weg 19, 01069 Dresden, Germany}
 \author{Juliane Simmchen}
 \affiliation{Pure and applied chemistry, University of Strathclyde, Cathedral Street, Glasgow, UK}
\author{Antonio Stocco$^*$}%
\affiliation{Institut Charles Sadron, CNRS UPR-22, 23 rue du Loess, Strasbourg, France}

\date{\today}% It is always \today, today, but any date may be explicitly specified

\begin{abstract}

The ability to design artificial micro/nanomachines able to perform sophisticated tasks crucially depends on the understanding of their interaction with biosystems and their compatibility with the biological environment. Here, Janus colloids fuelled only by glucose and light were designed, which can autonomously interact with cell-like compartments and trigger endocytosis.
The crucial role played by the far field hydrodynamic interaction arising from the puller/pusher swimming mode and adhesion is evidenced. It is shown that a large contact time between the active particle and the lipid membrane is required to observe the engulfment of a particle inside a floppy giant lipid vesicle. Active Janus colloids showing relatively small velocities and a puller type swimming mode are able to target giant vesicles, deform their membranes and subsequently get stably engulfed. An instability arising from the unbound membrane segment is responsible for the transition between partial and complete stable engulfment. These experiments shed light on the physical criteria required for autonomous active particle engulfment in giant vesicles, which can serve as general principles in disciplines ranging from drug delivery and microbial infection to nanomedecine.\\


\end{abstract}

\maketitle

\section{Introduction}

Catalytic microswimmers are micron-sized objects, that can self-propel and perform a variety of tasks on the micro-scale including drug delivery and environmental remediation. They achieve self-propulsion through chemical reactions that create concentration gradients that induce fluid flows in the surrounding solution. 
Efficient propulsion of the microswimmer requires a high reaction rate, which is usually achieved by the addition of highly reactive compounds such as hydrogen peroxide (\ce{H2O2}) \cite{Howse2007_H2O2} or hydrazine (\ce{N2H4}) \cite{Gao2014_hydrazine}, which are decomposed by the microswimmer. However, these toxic compounds limit potential applications e.g. in biological systems. 
Therefore, several works have focused on extending the range of propulsion reactions for catalytic microswimmers, including galvanic exchange reactions \cite{Feuerstein2021_galvanic}, photodeposition \cite{Wittmann2023_photodeposition}, the degradation of organic pollutants \cite{Wu2017_degradation,Zhang2017_degradation} or the oxidation of amines to imines \cite{Wittmann2022_amine}.\\ 
One very promising approach for powering of catalytic microswimmers in biological media is the use of a fuel that is also used by nature - glucose. \citeauthor{Wang2019_glucose} have shown, that \ce{Cu2O} microspheres can move in aqueous solutions of glucose under irradiation with visible light. The motion was explained with a photocatalytic oxidation of glucose to formic acid and the corresponding shorter monosaccharides arabinose, erythrose and glyceraldehyde \cite{Wang2019_glucose}. In their work, \citeauthor{Wang2019_glucose} used symmetrical spherical particles, and the symmetry breaking necessary for propulsion was introduced by shining light on one side of the particle. Due to self-shadowing effects, asymmetric reaction rates would indeed result in a directed motion away from the light source.
Here, we introduce Cu@\ce{SiO2} Janus particles that can move in an aqueous glucose solution under visible light irradiation. The Cu is converted to photocatalytic copper oxides, when the particles are dispersed in water. Due to their asymmetric Janus structure, they have an intrinsic ability to move in any direction independent of the light source, allowing us to observe more complex biologically relevant interactions with the surroundings.
In particular, interactions with fluid interfaces such as phospholipd membranes of cells represent a situation of confinement encountered in many processes. It was found that Janus swimmers are able to deform giant lipid vesicles when enclosed inside them \cite{vutukuri2020active}, a response that has also been confirmed for biological active matter \cite{le2022encapsulated} and by theoretical considerations \cite{lee2023complex, iyer2022non}.
Freely swimming Janus particles however, showed interactions that resemble the generic interactions of active matter with obstacles \cite{makarchuk2019enhanced, Simmchen2016, van2023transition}. For pusher type swimmers, persistent orbital motion of Janus swimmers around giant vesicles was reported \cite{Sharma2021}, and active transportation of the vesicles by a single particle could be achieved after applying strong forces \cite{Sharma2022}. To the best of our knowledge, no other behavior was reported up to now upon interaction of freely swimming active Janus particles with cell-like compartments.
While in previous experiments the Janus particles were usually made from Pt@\ce{SiO2}, which swim away from the metal cap and generally are considered to be pushers \cite{campbell2019experimental}, we use a new combination of Cu@\ce{SiO2} particles with glucose fuel. Similar to differently fuelled copper Janus particles \cite{Xiao2022, Sharan2022, Sharan2023}, these swimmers move towards the catalyst cap. This different swimming mode enables different types of interactions with vesicles, and allows to observe endocytic-like behaviors. 

\section{Results and discussion}

\subsection{Glucose fuelled Cu@\ce{SiO2} Janus colloid self-propulsion}
\label{sec1}

\begin{figure*}[t]
\includegraphics[width=1\linewidth]{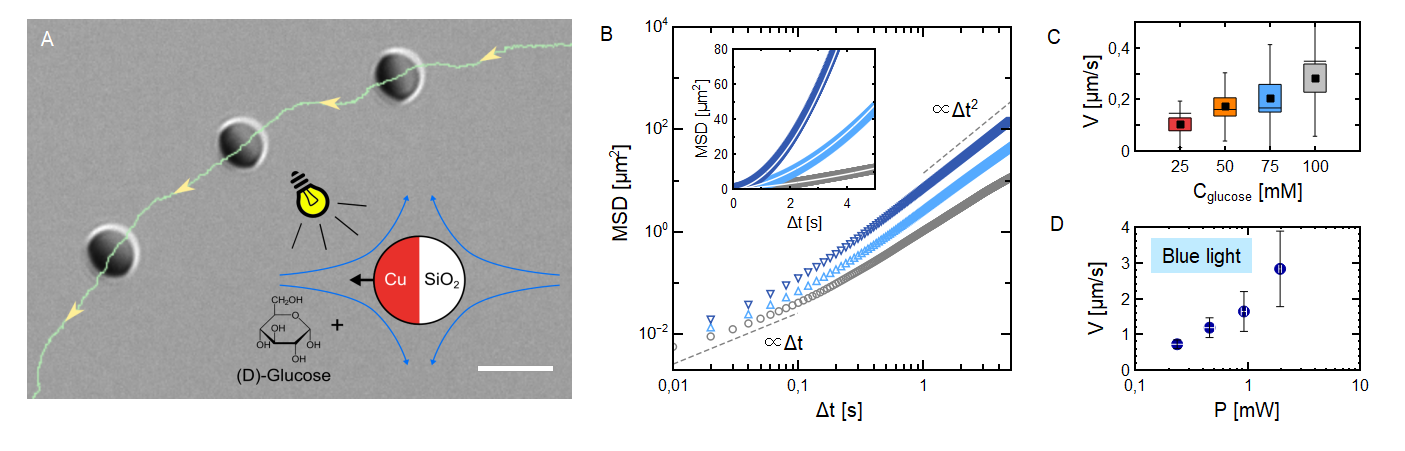}% Here is how to import EPS art
\caption{ (a) Bright field microscopy snapshot and trajectory (green) of a Cu@\ce{SiO2} Janus swimmer in a 100 mM glucose solution. Inset sketches the active motion mechanism. (b) Mean squared displacement as a function of lag time in log-log scale for three representative trajectories of particles immersed in 100 mM glucose with only white light (grey circles), white light and 0.46 mW blue light radiation (light blue up-pointing triangles) and white light and 1.96 mW blue light radiation (dark blue down-pointing triangles). Inset shows a linear scale representation with fits (white plain lines) allowing to extract projected velocity using equation 1. For these three cases, fit yields $V= 0.48 \ \mu$m.s$^{-1}$, $V= 1.23 \ \mu$m.s$^{-1}$ and $V= 2.33 \ \mu$m.s$^{-1}$ (bottom to top) (c) Projected active velocity $V$ extracted from MSD fits for particles immersed in glucose solutions of increasing concentration in normal illumination conditions. Black squares stand for averages on 10 trajectories. (d) Projected active velocity measured for particles immersed in a 100 mM glucose solution under different blue light illumination powers.}
\label{fig1}
\end{figure*}

Cu@\ce{SiO2} Janus particles were fabricated by thermal deposition of a 30~nm metallic copper layer on a monolayer of commercial silica particles of radius $R_P \approx 1.5 \ \mu$m, covering one hemisphere. These particles can self-propel catalytically when immersed in \ce{H2O2} and swim towards the metal cap \cite{Sharan2022}, which can be interpreted as behaviour as puller type swimmers because we assume the strongest slip is on the lead face \cite{Lauga_2020}. Here, we observe self-propelled motion in a different fuel. When immersed in a glucose aqueous solution at physiological concentration (25-100 mM), under normal microscope illumination, these particles show active propulsion as shown in Figure \ref{fig1}A. The sedimented particles navigate in a two dimensional plane close to the glass substrate, which allows to image them with bright field microscopy and track their center of mass with image treatment techniques (Supporting Information). During some segments of the trajectory, the plane defined by the Janus boundary is orthogonal to the imaging plane. The hemisphere of the particle  coated with the copper layer appears darker than the bare hemisphere and allows to track the in-plane orientation of the particle (Supporting Information). The strong correlation between the particle orientation and its direction of motion, after filtering out the Brownian diffusion, confirms the existence of a directional active propulsion (Supporting Information). The out-of-equilibrium nature of Cu@\ce{SiO2} dynamics in glucose can be assessed, and their projected active velocity quantified, by plotting the center-of-mass mean squared displacement (MSD = $\langle \left(x_{t+\Delta t}-x_t\right)^2 \rangle + \langle \left(y_{t+\Delta t}-y_t\right)^2 \rangle$) for single particles. For an active Brownian particle and for times shorter than the rotational diffusion time ($\tau_{ro}=1/D_{ro}= 8\pi \eta R_P^3/k_BT$ = 21 s) reads \cite{Howse2007_H2O2}:

\begin{equation}
\mathrm{MSD}\left(\Delta t\right)=
%\langle \left(x_{t+\Delta t}-x_t\right)^2 \rangle + \langle \left(y_{t+\Delta t}-y_t\right)^2 \rangle=
4D_{tr}\Delta t + V^2 \Delta t ^2,
\label{eq1}
\end{equation}

where $\Delta t$ is the time lag, $D_{tr}$ the transnational diffusion coefficient of the spherical particle (in the bulk $D_{tr,b}= k_BT/6\pi \eta R_P $) and $V$ an average projected speed on the plane of observation. The MSD of Cu@\ce{SiO2} particles in glucose, shown for representative cases in Figure \ref{fig1}B, clearly displays a linear (diffusive) regime followed by a quadratic (ballistic) regime at longer times. The crossover between those two regimes occurs at a characteristic timescale $\tau_c=4D_{tr}/V^2$ where the passive and active contributions are comparable.
Fitting MSD data with Equation (\ref{eq1}) allows to extract both  the translational diffusion coefficient and the projected active velocity for single trajectories (fits are plain white lines in the inset of Figure \ref{fig1}B). We can show that an increase of the glucose concentration results in an increase of the projected speed, see Figure \ref{fig1}C. Similarly, illuminating the sample with blue light of increasing power will also enhance the measured projected velocities as shown in Figure \ref{fig1}D. We have also evaluated the effect of the white light illumination on the active speed (Supporting Information). Values of $D_{tr}$ extracted from MSD fits for lower glucose concentrations (i.e. low activity) agree with the diffusion of particles close to a solid wall for a gap distance $h \approx 300$ nm (Supporting Information). For higher glucose concentrations, $D_{tr}$ increases and becomes comparable to the bulk prediction. Either an increase of the gap distance between the particle and the substrate or additional noise, both induced by the active propulsion mechanism can explain this increase of the the measured $D_{tr}$. For pusher swimmers however, it was reported that the gap distance to the substrate remains constant for increasing degrees of activity \cite{Ketzetzi2020}.

\subsection{Particle-vesicle encounter and engulfment phenomenon}
\label{sec2}

\begin{figure*}[t]
\includegraphics[width=1\linewidth]{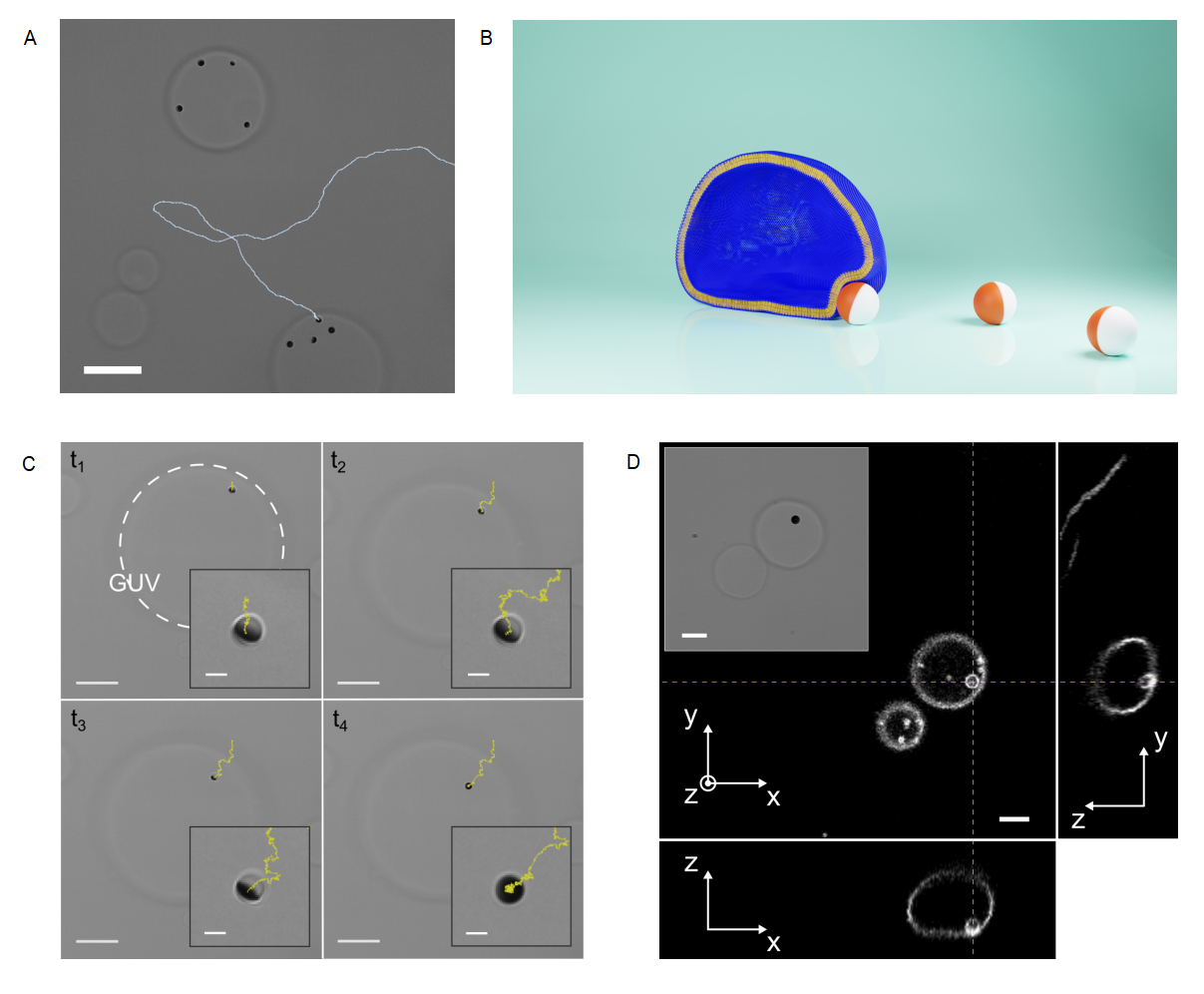}% Here is how to import EPS art
\caption{(a) Bright field microscopy snapshot with the backward in time trajectory of a particle in contact with a POPC GUV. (b) Sketch of the situation in (a). (c) Snapshots of a bright field microscopy acquisition showing wrapping of a Janus Cu@\ce{SiO2} by a POPC GUV. Backward in time trajectory is shown in yellow. (d) Confocal microscopy images showing the xy, zy and zx planes allowing to show the fluorescence signal from the lipids corresponding to the GUV and the wrapped particle. Inset shows the corresponding bright field image. }
\label{fig2}
\end{figure*}

In this section, we illustrate the phenomenon observed when an active Cu@\ce{SiO2} particle in a glucose solution spontaneously interacts with a giant unilamellar vesicle (GUV). This phenomenon can be described as composed of two steps: (i) the hydrodynamic capture, and (ii) the particle being wrapped by the vesicle membrane. Note that the particle activity depends on exposure to visible light and physiological glucose concentrations. Hence, we got rid of the limitations of using \ce{H2O2} as a fuel, which is not biocompatible and generally harmful for cells \cite{BLOCK1997,VALEN19991480}. In all experiments, we used GUVs made of phospholipids in the low-tension membrane regime, which are pertinent systems to study endocytic-like phenomena \cite{Dai1997,PONTES201730} in the absence of receptor-mediated interaction or large external forces \cite{Meinel2014,Spanke2020,Fessler2023}. 
Figure \ref{fig2}A illustrates a typical result obtained few minutes after the spontaneous interaction between active particles and GUVs. Some active particles are already engulfed by GUVs and for one isolated particle we could track its active trajectory before and after encountering a GUV. The impact of the activity on the encounter rate is evident as it allows the particles to explore larger areas in a given time with respect to passive Brownian particles. The impact of far-field hydrodynamic interactions on the encounter rate is hard to evidence, because directional changes can be due either to hydrodynamic interaction or Brownian rotational diffusion. Still, even in diluted situations this kind of event happens at a high rate and after few minutes, most of the active particles in the sample are engulfed by GUVs (Figure \ref{fig2}). In the first step of the interaction,
 the particle is confined between the bottom wall and the GUV membrane, which is slightly deformed by the particle, as sketched in Figure \ref{fig2}B. The first two snapshots at times $t_1$ and $t_2$ in Figure \ref{fig2}C correspond to bright field microscopy images of the \textit{capture}. %when the particle is in contact with the GUV and deforms it
 After the hydrodynamic capture, which can last several tens of seconds, a transition occurs and a sudden radial displacement of the particle towards the projected GUV center of mass is observed at the time $t_3$ in Figure \ref{fig2}C (showing snapshots of the Movie S1, Supporting Information).
  At time $t_4$, the particle appears completely dark, evidencing a change of orientation and the trajectory shows a more confined diffusion than previously. Using confocal microscopy, we can show that at this final stage, the particle is inside the GUV volume while being fully wrapped by the lipid membrane (Figure \ref{fig2}D), and remains in this stable state indefinitely. Movie S2 (Supporting Information) shows the whole wrapping process in fluorescence microscopy, together with a particle that was already engulfed. 
  Note that in our experimental conditions, the wrapped particle is connected to the mother vesicle with a small neck or tube \cite{Fessler2023}.
  In the next sections, we will describe quantitatively the two different steps of the phenomenon introduced here.

\subsection{Active particle hydrodynamic docking at the vesicle periphery}
\label{sec3}

\begin{figure}[t]
\includegraphics[width=1\linewidth]{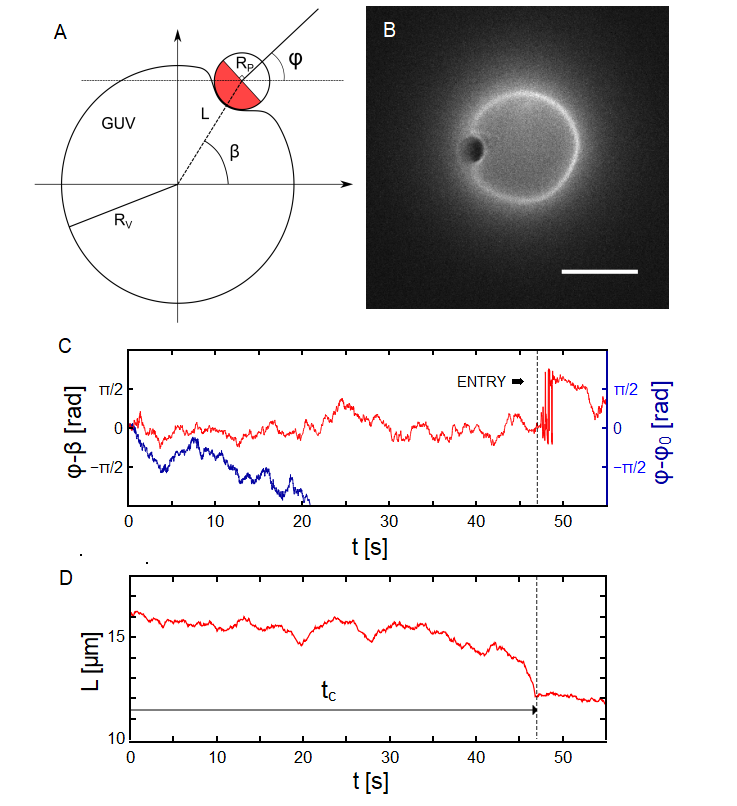}% Here is how to import EPS art
\caption{(a) Sketch defining the angles $\varphi$, $\beta$ and the distance $L$. (b) Fluorescence microscopy image of a particle in the capture phase, deforming the GUV. Scale bar is 10 $\mu$m (c) Red curve shows the temporal evolution of the angular difference $\varphi-\beta$ for a particle in contact with a GUV before being wrapped. The blue curve shows the evolution of the angle $\varphi$ with respect to $\varphi_0=\varphi(t=0)$ for a particle far from a GUV. (d) Evolution of the distance $L$ over time for a particle in contact with the GUV at $t=0$.}
\label{fig3}
\end{figure}

In order to elucidate on the mechanism responsible for capture of the particle at the GUV periphery, we quantify here the particle motion focusing our attention on the rotational dynamics. During the \textit{capture} phase, the 
%particle encountered a vesicle and is subsequently deforming it, with the 
copper cap of the particle is indeed systematically facing the GUV membrane. The situation is sketched in Figure \ref{fig3}A, together with the definition of the particle orientation angle: $\varphi$, the azimuthal angle of the particle radial position: $\beta$ and the distance between the particle and GUV centers: $L$. The active propulsion force is oriented towards the lipid membrane and promotes the deformation of the latter. This orientation can be observed in bright field microscopy as in Figure \ref{fig2}C and deformation of the vesicle can be confirmed using fluorescence microscopy as shown in Figure \ref{fig3}B.

The time evolution of the angular difference $\varphi - \beta $, plotted in Figure \ref{fig3}C, shows a  zero mean value $\langle \varphi -  \beta \rangle \approx 0$, and a clear orientational confinement of the particle  with the Copper cap facing towards the GUV projected center. The confinement effect is apparent when comparing $\varphi - \beta $ to the time evolution of the orientation $\varphi - \varphi_0 $ of an active particle far from the GUV, also shown in Figure \ref{fig3}C. 
Here, the particle orientation $\langle \varphi -  \beta \rangle \approx 0$ is very different from the one observed for pusher type swimmers, which align as $\langle \varphi -  \beta \rangle \approx \pi/2$ performing orbital motion around GUVs (Supporting Information) or solid spherical obstacles \cite{Sharma2021,Simmchen2016,Das2015,Spagnolie2012}. 
These results point to a self-propulsion mechanism of Cu@\ce{SiO2}colloids in glucose generating puller-type swimmer flow-fields, as it is in \ce{H2O2} with the Cu cap forward. Far-field hydrodynamics models indeed predict that puller type swimmers encountering a spherical obstacle should be trapped by the interface and remain motionless even for reasonably low dipole strengths and small obstacles \cite{Spagnolie2015}. Hence, the \textit{capture} observed in our experiments can be attributed to the hydrodynamic attraction expected for puller active particles close to obstacles. The additional complexity arising from the softness and fluidity of the GUV membrane does not seem to modify the attractive nature of those interactions.
The particle orientation is however not completely frozen and $\varphi -  \beta $ show significant fluctuations, which indicates rotational diffusion at short times while experiencing an effective restoring torque at long times. Hence, we calculated the mean squared angular displacement (MSAD) of $\varphi-\beta$. At short times, the slope of the MSAD allows to extract the in-plane rotational friction $\zeta_{\varphi}$ experienced by the particle during the capture phase. The fit yields $\zeta_{\varphi} = 7.3 \times 10^{-20}$ N.s.m. (Supporting Information) which is close to the theoretical value expected for the rotational friction experienced by a spherical $R_P= 1.5 \ \mu$m particle in the bulk $\zeta_{ro,b}=8\pi\eta R_P^3= 8.5 \times 10^{-20}$ N.s.m. 
Hence, no additional dissipations due to the GUV membrane were detected by the particle during the \textit{capture} that points to a large water gap between the membrane and the particle surface, which also agrees with a far-field hydrodynamic interaction. 

\begin{figure*}[t]
\includegraphics[width=1\linewidth]{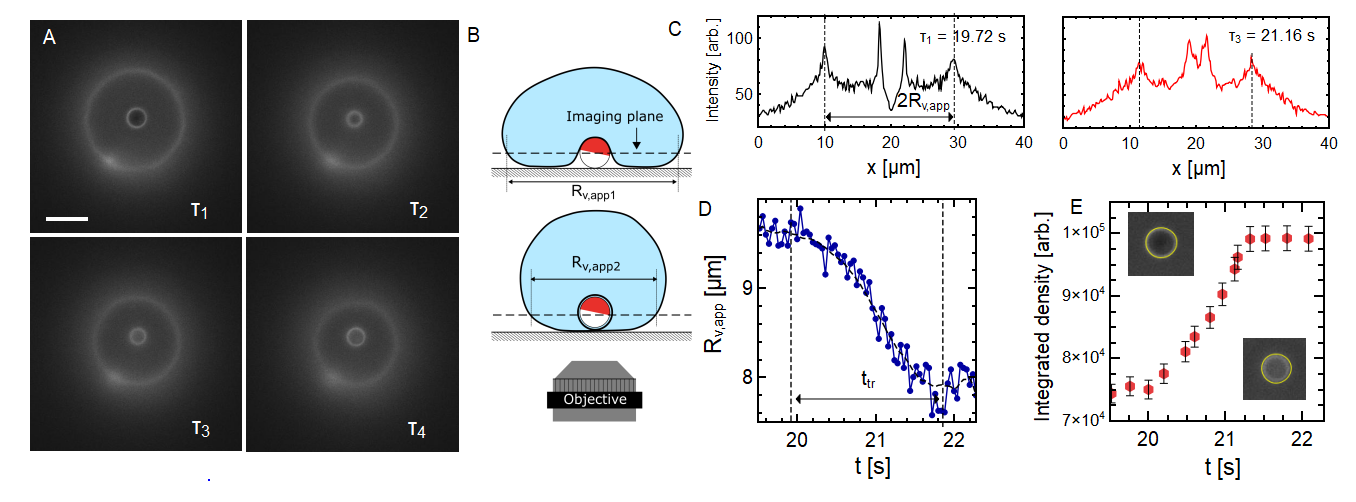}% Here is how to import EPS art
\caption{(a) Snapshots of a fluorescence microscopy acquisition showing the partial wrapping to full wrapping transition in the case of a GUV sitting on top of the particle. (b) Sketch illustrating the situation in (a). (c) Pixels intensity profile along an horizontal line passing by the center of the particle. (d) Evolution of $R_{v,app}$ measured from intensity profiles during the full wrapping transition. (e) Temporal evolution of the pixel integrated density (total intensity) in a circular region centered on the particle during the wrapping transition.}
\label{fig4}
\end{figure*}

The magnitude of the MSAD($\varphi-\beta$) plateau at long times provides an angular range $ \varphi ^* = \sqrt{\langle \left(\varphi -  \beta\right) ^2\rangle _{sat}} $  $\approx 0.47$ rad, which describes the competition between the Brownian diffusion that tends to randomize the particle orientation and an effective restoring torque tending to align the particle towards $\varphi-\beta$ = 0. As described before, the origin of this restoring torque can be purely hydrodynamic. In this first stage of the interaction between the active colloid and the GUV characterized by far field hydrodynamic effects, the adhesion energies between the two particle (Copper and Silica) faces and the GUV membrane are not expected to play a significant role for the particle orientational confinement.

The transition to full wrapping is evidenced by an abrupt variation of the distance $L$ in Figure \ref{fig3}D at the end of the capture phase. The approaching of the particle to the GUV center of mass is however not always a direct effect of the transition to full wrapping. It happens that the particle first migrates towards the GUV projected center of mass and remains confined between the GUV and the substrate before undergoing the full wrapping transition. It can also be seen in Figure \ref{fig3}C that after the transition (represented by the vertical dashed line), the in-plane orientation $\varphi$ can not be precisely measured and show large variations as a result of the out-of-plane reorientation of the particle, which does not allow to image the Janus boundaries. 
During the capture step, there is a dynamics associated with the drainage of the water film such that  the particle-membrane distance is not the equilibrium one. Indeed, the particle-membrane surface interaction potential, in the case of non-specific adhesion, is minimal at separation distances in the range 20-100 nm \cite{Radier1995,Cantat2003} while the basin of attraction associated to the hydrodynamic capture is of the order of the swimmer body size (few micrometers here) \cite{Spagnolie2015}. The characteristic time $\tau_d$ associated to the drainage of the water film can be estimated from Reynolds law \cite{Bernard2000} and yields $\tau_d \approx 10^1$ s with the active propulsion as driving force. To investigate the dynamics towards the wrapping transition, measurements on the membrane profile upon wrapping and the driving forces of the process will be discussed in the following section.

\subsection{Wrapping transition dynamics}

The dynamics of the GUV wrapping the active particle can be followed using fluorescence microscopy. Taking advantage of a peculiar situation that can be encountered (depicted in Figure \ref{fig4}A and B) where the GUV is located exactly on top of the particle, one can monitor the evolution of the projected GUV area and particle fluorescence during a wrapping dynamics. Indeed, the geometry does not allow to image the deformations of the GUV induced by the particle as it can be the case for particle wrapping induced by optical tweezers \cite{Fessler2023} occurring in the equatorial plane of the GUV. In Figure \ref{fig4}A, one can see fluorescence microscopy snapshots of an active particle that came in contact with the GUV, and instead of remaining at the periphery as in the cases described before, slowly migrated under the GUV until sitting at the center of the circular projected GUV area. At time $\tau_1 =19.72 $ s after the start of the acquisition, the particle is not wrapped yet and the situation corresponds to the one depicted on the top panel of Figure \ref{fig4}B. At time $\tau_4 = 22.08$ s, the particle is fully wrapped and the situation corresponds to the situation of the lower panel in Figure \ref{fig4}B, while intermediate times correspond to the dynamics of full wrapping transition ($\tau_2= 21.12$ s and $\tau_3= 21.16$ s). In Figure \ref{fig4}C, we show representative intensity profiles passing through the projected GUV center of fluorescence microscopy images from Figure \ref{fig4}A. The two external peaks correspond to the bright circle coming from the GUV projection in the imaging plane. Such profiles allow to follow the evolution of an apparent projected radius $R_{v,app}(t)$ during the full wrapping transition from the distance between these two peaks.

Plotting the evolution of $R_{v,app}$ during the transition (from $\tau_1$ to $\tau_4$) in Figure \ref{fig4}D shows a decrease of more than a micron, which is a signature of the full wrapping transition. Indeed, fully enveloping the particle with membrane requires pulling more membrane surface area to the wrapping site resulting in a decrease of the apparent vesicle projected radius (as depicted in Figure \ref{fig4}B) as the GUV sphericity increases. Conversely, the fluorescence signal emitted from the region corresponding to the particle projected area (quantified by the so-called pixel integrated density in Figure \ref{fig4}E) increases following the same dynamics, confirming the full wrapping of the particle. During this transition, the contact line advances from the equator of particle (the Janus boundary) to its pole. We show here that this transition occurs in a time $t_{in} \approx 2$ s, resulting in a contact line velocity $v_{c}= \pi R_P/2 t_{in}= 1.17 \ \mu$m.s$^{-1}$. This is comparable to contact line velocities measured in the system of Spanke et al. \cite{Spanke2022}, where wrapping was triggered by depletion attraction with adhesion energy densities of the order of 10$^{-6}$ N.m$^{-1}$ for similar particle size ($R_P = 1 \ \mu$m).

\begin{figure}[t]
\includegraphics[width=1\linewidth]{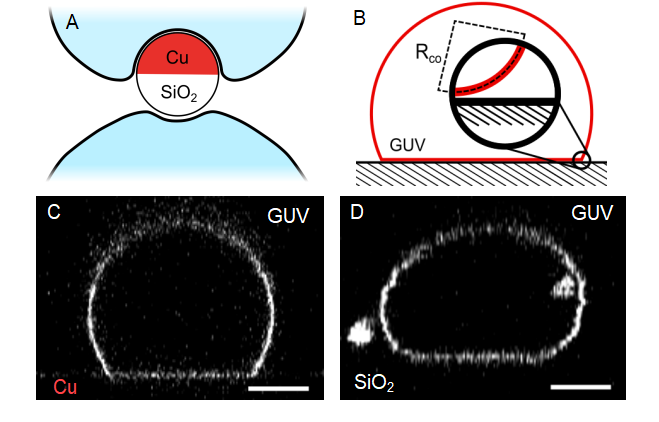}% Here is how to import EPS art
\caption{(a) Sketch illustrating the difference of GUV membrane interaction with the copper coated surface and the bare \ce{SiO2} surface of the Janus particle. (b) Sketch of the experiment and definition of the contact curvature radius $R_{co}$. (c) Representative transverse views (xz plane) from a confocal microscopy acquisitions of osmotically deflated POPC GUVs sedimented on a Copper coated glass substrate and (d) on a bare glass substrate.}
\label{fig3b}
\end{figure}

In order to evaluate the adhesion energy densities in our system,
we performed adhesion experiments of GUVs on planar Silica and Copper surfaces \cite{Lipowsky1991,Gruhn2007,Bernard2000}. We are able to thermally deposit a nanometric Copper layer on glass slides over several $cm^2$ surface area using the same technique employed to deposit the cap on the Silica particles. Figure \ref{fig3b}A shows the analogy between our situation and the experiment we performed to determine $w$ illustrated in Figure \ref{fig3b}B, where we also define the quantity $R_{co}$ which will be of interest here. Using a confocal microscope, we can extract the full 3-dimensional shape of the adhesive vesicles. If we compare typical side view profiles (x-z plane) acquired with confocal microscopy of an initially floppy vesicle on Copper (Figure \ref{fig3b}C) or on Silica glass (Figure \ref{fig3b}D), it is clear that the interaction with the bottom substrate is very different leading to different vesicle shapes. The shape of the unbound part of the membrane is very close to a spherical cap (expected in the limit of high adhesion \cite{Bernard2000}) in the case of Cu substrate, while the shape is more elongated and the membrane appears more fluctuating for Silica. Indeed, the bending energy that tends to maximize the membrane curvature over the whole surface area is in competition with the adhesion energy that tends to maximize the membrane area in contact with the surface. The curvature radius at contact therefore reads \cite{Lipowsky1991}:
\begin{equation}
R_{co}= \sqrt{\frac{\kappa_b}{2w}},
\label{eq2}
\end{equation}
with $\kappa_b$ the bending modulus of the membrane. For a deflated GUV on a Copper covered glass slide, the resolution of the profile acquired with confocal microscopy does not allow to fit a circle to extract a curvature. The fact that the curvature is beyond the limit set by our resolution allows however to set a lower bound for the curvature which yields a lower bound for the adhesion energy density $w$. If one takes $R_{co,Cu} \le 0.5 \ \rm{\mu}$m as the resolution-limited higher bound of measurable curvature radius, we get $w_{Cu} \ge 2 \times 10^{-7}$ N.m$^{-1}$ as a bounding value for the adhesion energy density. For the case of the bare glass substrate, however, the situation is different and one can always define a radius of contact curvature using confocal images and   ImageJ analysis techniques \cite{STEINKUHLER20161454}. The average fitted radius on $N=$ 5 different GUVS is $\langle R_{co,SiO_2}\rangle =7.5 \pm 0.6 \ \mu$m yielding $w_{SiO_2} = 8.8 \pm 1.3\  \times 10^{-10}$  N.m${^-1}$.

Hence, the associated adhesion energy density $w_{Cu}$ is several orders of magnitude larger than $w_{SiO_2}$ in this geometry. The origin of this adhesion can be attributed to electrostatic attraction between the positive zeta potential of Copper \cite{Sharan2022,Sharan2023} and the slight negative zeta potential of POPC vesicles measured for low electrolyte concentrations \cite{POPCcharge}. While it was shown that Cu$^{\rm{2+}}$ ions can bind to PC and PE lipid headgroups \cite{Jiang2018,Poyton2016} and induce structural changes on the membrane, it is not obvious whether these effects can be responsible for an effective attraction between the two surfaces. Still, such adhesive behaviors are consistent with our measurements and the strong reduction of the translational diffusion constant $D_{tr}$ measured once the particle is fully wrapped ($D_{tr}\approx D_{tr,b}/3$, Supporting Information) highly suggests a small particle-membrane water gap thickness. Indeed, this large drag increase accounts for the dissipations associated to the membrane neck diffusing in the plane of the lipid membrane.

\subsection{Theoretical modelling of the wrapping energy landscape}

In the previous sections, we showed that active  Cu@\ce{SiO2} Janus particles are able to deform GUV membranes until reaching a stable complete engulfment. We also measured the adhesion energy densities and the characteristic capture time. By using these experimental values, in this section we model the driving energies and costs associated to the particle engulfment, in order to understand its stability. Our model will also allow to address questions regarding the dynamics of the phenomenon, in particular the duration of the capture time $t_c$ before the wrapping transition. %To do so, one has to consider the system as a whole and consider the energies associated to the involved deformation modes of the membrane together with adhesion and active propulsion force contributions.

\begin{figure*}[t]
\includegraphics[width=1\linewidth]{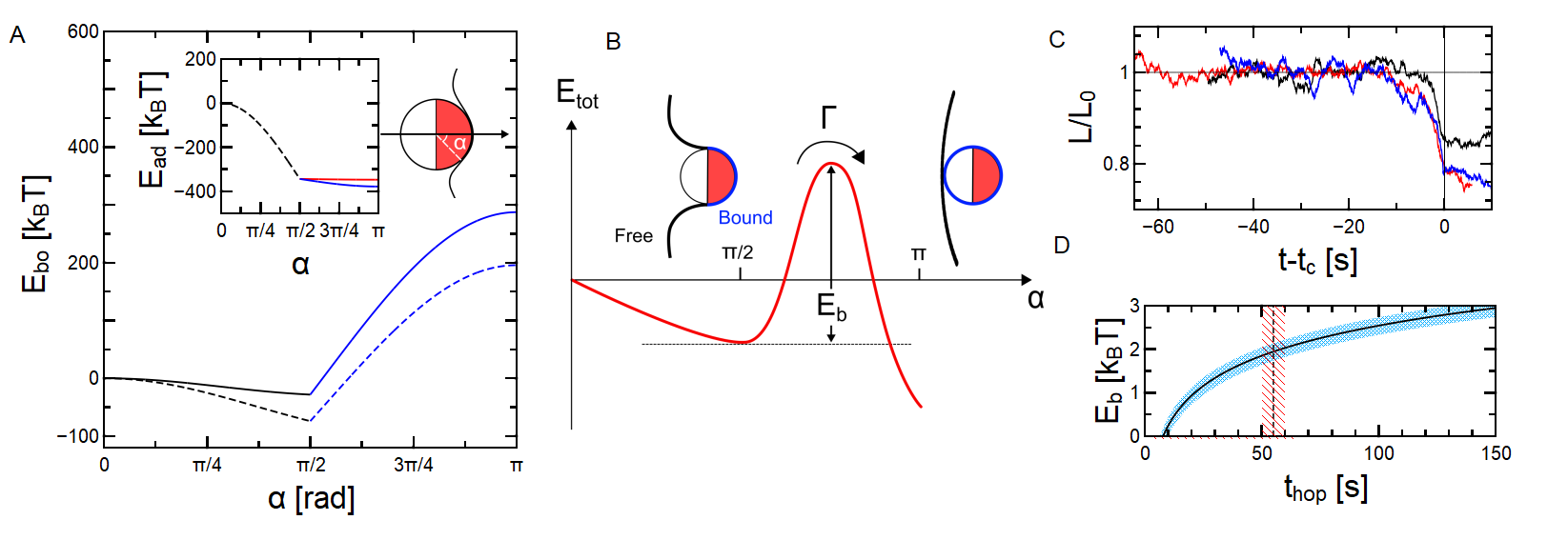}% Here is how to import EPS art
\caption{(a) Energy associated to the membrane segment bound to the particle $E_{bo}$ as a function of wrapping angle. Plain line stands for spontaneous curvature $m=0$ and dashed one to $m=-10^5\ \mu$m$^{-1}$. Insets show the adhesion energy between the membrane and the particle as a function of wrapping angle considering the particle is wrapped with the copper facing towards the membrane. (b) Illustration of the shape of the total energy ($E_{bo}+E_{free}$) and definition of the energy barrier $E_b$ between the partial and full wrapping states. (c) Temporal evolution of $L/L_0$ (where $L_0=L(t=0)$) for three wrapping experiments with $t_0$ taken as the moment the particle undergoes the full wrapping transition. (d) Height of the energy barrier $E_b$ as a function of the inverse hopping rate $1/\Gamma=t_{hop}$ using the Kramers theory \cite{Kramers,Brown2016} of escape over an energy barrier using stiffness and diffusivity extracted from curves in (c). Vertical dashed line stands for average $t_c$ from (c).}
\label{fig5}
\end{figure*}

The system considered in our model is composed of a particle and a fluid membrane with bending rigidity $\kappa_b$, tension $\sigma$ and particle-membrane adhesive energy per unit area $w$. If one assumes a flat membrane and negligible volume conservation constraints, the free energy of the system can be described with the following Helfrich-Canham Hamiltonian \cite{Wu2022}:

\begin{equation}
E_{\text{tot}} =  \int_{A_{tot}}^{} \frac{\kappa_b}{2}(2M-m)^2 \,dA - \int_{A_{b}}^{} w \,dA + \sigma\Delta A - fh
\end{equation}

with $M$ the membrane mean curvature, $m$ a spontaneous membrane curvature, $\Delta A$ the membrane excess area pulled to the wrapping site, $f$ the propulsion force and $h$ penetration depth. Note that the Helfrich Hamiltonian usually contains a Gaussian curvature term, which will be ignored in the following due to the absence of topology change \cite{Deserno2004}. Let us first consider that the adhesion, bending and tension contributions of the membrane segment \textit{bound} to the particle dictates the wrapping degree $\alpha$. Thus, we ignore at the moment the contribution of the free segment (the vesicle area that is not bound to the particle). In the case of a spherical particle with radius $R_P$, the energies associated with tension and bending for the bound membrane area $A_b= 2\pi R_P^2(1-\cos \alpha)$ are:  $E_{\sigma}= \pi \sigma R_P^2(1-\cos \alpha)^2$ and $E_{b} = 4\pi \kappa_b (1+ mR_P) (1- \cos \alpha)$ respectively, where $m$ is a membrane spontaneous curvature. These two contributions can be calculated in our system using a standard value of the bending rigidity $\kappa_b=27.3 \pm 5.1 \ k_BT\approx 10^{-19} $ J for POPC vesicles in glucose/sucrose solutions \cite{Faizi2022} and a tension $\sigma= 10^{-8}$ N/m as an average accounting for the tension distribution from previous force measurements on floppy vesicles prepared with the same protocol \cite{Fessler2023}. 
The two negative (energetically favorable) contributions to the free energy of the system can also be estimated. The work done by the propulsion force $f$ can be expressed as a function of $\alpha$ as $E_f=-fR_P(1-$cos$\alpha)$. Here, an upper bound of the active force $f$ in our system can be calculated by taking it to be equal to the transnational friction force when the particle is far from a GUV at the highest measured projected velocity $V$. We then have $f=\zeta_{tr,b}V=6\pi \eta R_P V \approx 0.1 $ pN. Considering the Janus geometry of the particle, the adhesion energy density $w$ is not constant throughout the wrapping process (as shown previously). Indeed, the adhesion energy density of Copper $w_{cu}$ and Silica $w_{SiO_2}$ with the membrane are very different, as shown in the previous section. The adhesion energy then reads:

\begin{equation}
  \begin{aligned}
     E_{ad} & =
    \begin{cases}
      2\pi R_P^2\left(1-\cos\alpha\right)w_{Cu} &  \alpha < \pi/2 \\
      2\pi R_P^2w_{Cu}+2R_P^2\left(\sin \left(\alpha-\frac{\pi}{2}\right)\right) w_{Si}&  \alpha \ge \pi/2\\
    \end{cases}    \\
  \end{aligned}
\label{eqad}
\end{equation}

Taking $w_{Cu} = 0.2 \ \mu$N.m$^{-1}$ the lower bound inferred from confocal microscopy acquisitions and $w_{Si} = 0.88 $ nN.m$^{-1}$ allows to have a full description of the wrapping energy landscape for the bound membrane segment. The unbound membrane segment is usually considered to be of negligible importance in equilibrium systems in the absence of spontaneous curvature as it is expected to adopt a minimal surface shape (close to a catenoid) minimizing the mean curvature $M$. In Figure \ref{fig5}A we plot the sum of the contributions for the bound membrane segment together with the active force and adhesion contributions $E_{bo} = E_b+E_{\sigma}+E_f+E_{ad}$ as a function of wrapping angle $\alpha$ defined in the sketch. Inset shows the evolution of $E_{ad}$ alone when taking into account the Janus geometry (\ref{eqad}). It appears clearly that using values of $w_{Cu}$ inferred from experiments, the adhesion is what drives the wrapping and not the active force as it is much weaker, $|E_{ad}| > |E_{f}|$. However, due to the Janus geometry and the difference in the Cu and \ce{SiO2} adhesion energy density, a stable equilibrium minimum is predicted at $\alpha=\pi/2$ for $E_{bo}$. Note that the evolution of $E_{bo}(\alpha)$ is qualitatively the same in the presence of a negative spontaneous curvature $m=-10^5$ m$^{-1}$ expected in our system due to the inner sucrose/outer glucose solution asymmetry across the membrane \cite{Fessler2023,Gunther1999}. Therefore a stable state should correspond to a particle partially wrapped up to its equator, which is not consistent with our experiments at long times; but it may describe the transitory state observed experimentally in Figure \ref{fig3}B. Indeed, we showed in Figure \ref{fig4} that a particle in that state can undergo a transition to full wrapping after some time (i.e. a transition to $\alpha=\pi$). However, $E_{bo}$ at $\alpha=\pi$ is very high and energetically unfavorable. Therefore, a model taking into account only the $\it{bound}$ membrane segment can not describe our results. Hence, we consider a new scenario accounting for a significant contribution of the energy associated to the shape of the free membrane segment, $E_{free}$. In Figure \ref{fig5}B, we sketched an energy profile that can describe the transition to full wrapping (as in Figure \ref{fig4}) with $E_{tot}(\alpha)=E_{bo}(\alpha)+E_{free}(\alpha)$.
The unbound (free) membrane segment may not adopt minimal energy shapes and have significant energetic cost associated to it for $0<\alpha<\pi$. When $\alpha=\pi$ (full wrapping), a small neck is formed behind the particle and the free part of the membrane adopts a shape with $M\approx 0$. One might think that the neck, as a highly curved segment is very costly in terms of bending energy. It appears that in the contrary, this structure is very stable \cite{Fessler2023} as this high curvature occurs on a very small scale involving few lipids which is beyond the curvature model for lipid membranes and  acts more like a defect or singularity \cite{Deserno2004}. There therefore exists an energy barrier $E_b$ that separates the metastable partial wrapping state and the stable full wrapping state. Modelling this energy barrier would involve calculating the energy $E_{free}$ which is only possible if one knows the shape adopted by the free membrane. The geometry here does not allow to extract membrane profiles from fluorescent microscopy images. However, for cases when the capture time is large compared to the fast local relaxation processes in the potential well ($\sim$ 1 s ) around $\alpha=\pi/2$ (such as for the cases in Figure \ref{fig5}C), we can use the Kramers theory of escape over an energy barrier. The latter links a hopping rate $\Gamma$, i.e. the inverse average time spent by the particle in the metastable state before hopping $\langle t_{hop}\rangle $, to $E_b$ such as \cite{Kramers,Brown2016}:  

\begin{equation}
\Gamma=\frac{1}{\langle t_{hop} \rangle} =  \frac{k D_{\alpha}}{2\pi k_B T} \exp\left( - \frac{E_b}{k_B T}\right)
\label{eqkramers}
\end{equation}

where $D_{\alpha}$ characterizes the dynamics of the fluctuations of the wrapping angle $\alpha$ in the potential well of stiffness $k$ (in the harmonic approximation) around the minimum $\alpha=\pi/2$. Note that $D_{\alpha}$ has the dimension of a rotational diffusion coefficient (without being one) and $k$ is an energy. Extracting $k$ and $D_{\alpha}$ in our system would imply being able to resolve the fluctuations of $\alpha$ in the partial wrapping state, which is not possible with our setup. One quantity that we can measure however is the distance $L$ between the particle and the GUV center of mass, see Figure \ref{fig5}C. We can impute the fluctuations of $L$ to fluctuations of $\alpha$ as:

\begin{equation}
\Delta L = R_P \left(\sin \left(\frac{\pi}{2}-\alpha \right) - \sin \left(\frac{\pi}{2}-(\alpha+\Delta \alpha)\right)\right)
\end{equation}

Which in the small angle approximation $\sin(x)\approx x$ (as $\alpha\approx \pi/2$ in the half wrapped state and $\Delta \alpha \ll 1$), yields:

\begin{equation}
\Delta L = - R_P\Delta \alpha
\end{equation}

Hence, we can plot the distribution of $\alpha$ inferred from $L$ and translate it in an effective potential $\Delta U (\alpha)$, which we can fit assuming it is quadratic around $\alpha=\pi/2$ to extract the stiffness $k$. The average stiffness of the quadratic trapping potential using our data is $k=38 \pm 4 \ k_BT$ (Supporting Information). Diffusivity $D_{\alpha}$ can be inferred from a linear fit at short times of the mean squared angular displacement curves of $\alpha$ as a function of lag time. The mean diffusivity $D_{\alpha}$ measured here is $D_{\alpha}=0.021 \pm 0.002$ rad$^2$.s$^{-1}$ (Supporting Information). We can plot $E_b$ as a function of $t_{hop}$ (Figure \ref{fig5}D) by reversing Equation \ref{eqkramers} and by plugging in the calculated $k$ and $D_{\alpha}$. This allows to visualize the expected magnitude of $E_b$ for hopping times few tens of seconds long which is what we observe in experiments as a lower bound. Note that $t_{c}$ can be larger than $t_{hop}$ as it might take some time for the water gap between the particle and the membrane to drain and the adhesion measured in equilibrium to be effective. Still, $t_{c}$ provides an upper bound for $t_{hop}$ and this reasoning allows to estimate that the full energy profile considering the full (bound and free) membrane shape leads to an energy barrier $E_b\approx 2\  k_B T$, see Figure \ref{fig5}D. 

\section{Conclusion}

To conclude, we designed a system that autonomously reproduces endocytosis and the wrapping of artificial active swimmers by giant vesicles, only by providing visible light and physiological glucose concentrations. We showed that the velocity of the active particle is not the quantity to maximize in order to observe engulfment by cell-like compartments. In fact, our results evidence the key role of long contact times between the swimmer and the wrapping host. Long contact times during which the active propulsion force of the particle is directed towards the membrane indeed allow the liquid film between the two objects to drain and physicochemical interactions to be triggered leading to adhesion between the two surfaces. Additionally, time is needed for the system to overcome energy barriers separating two wrapping states. It is therefore crucial that far-field interactions, such as the long-range hydrodynamic attraction between the puller swimmer and the obstacle here, prevent the particle to swim away from the vesicle membrane. 
To generalize, we can predict for future investigations that bigger particles sizes would facilitate the hydrodynamic docking of the particle at the vesicle periphery, due to reduced magnitude of the orientational and translational Brownian noise. However, the associated increase in tension cost will lead to a higher energy barrier and therefore increase the average hopping time over the barrier. This leads to a more effective but slower overall process. For smaller particles, the decrease of the tension energetic cost would be compensated by an increase in bending energy cost, leading to a similar barrier height. The enhanced translational and rotational diffusion would however prevent the slow processes to occur as the particle is more likely to diffuse away or reorient before adhesion is triggered. Note that from Reynold's law of drainage, we expect the duration of the film drainage step to be larger for larger spherical particles. However, a non-spherical particle could minimize the drainage time to trigger adhesion faster.\\
Overall, this research provides insights on the physical criteria required to trigger autonomous full and stable wrapping of active particles, when their direction of motion is not prescribed by an external field. Such findings are of particular interest for interactions of swimmers in general with biological units to perform tasks at the cellular level in contexts such as drug delivery and nanomedicine, or in the case of microbial infections. In the scope of micromotors design for wastewater treatments, being able to get all particles stably wrapped by fluid membranes once they performed a given task could constitute an efficient way to remove them from any environment, by removing the membrane enclosing them as a whole instead of individually removing particles.

\section{Experimental section}

\textit{Janus particles.} Janus colloids were fabricated by drop casting a diluted solution of silica microspheres of radius $R_P=1.5\ \mu$m (Sigma Aldrich) on a plasma clean glass slide. After evaporation of the solvent, a nanometric (30 nm) Copper layer was subsequently thermally deposited on the microspheres monolayer. The half Copper-coated microspheres were released in MilliQ water thanks to ultrasonic pulses and stored at 4$^{\circ}$C until use.

\textit{Giant unilamellar vesicles assembly.} The giant unilamellar vesicles were prepared using a PVA (Polyvinyl alcohol) gel-assisted formation method \cite{Weinberger2013}. The PVA gel is prepared by dissolving PVA in pure water (MilliQ water) at 5 $\%$ w/v concentration. The prepared PVA gel is spread uniformly in PTFE (Polytetrafluoroethylene) troughs and dried for 45 minutes at 80°C in an oven. In the case of POPC/POPC-NBD vesicles, 5 µL of a 99:1 (molar) mixture of POPC (1-Palmitoyl-2-oleoylphosphatidylcholine) and POPC-NBD (POPC fluorescently labelled with Nitrobenzoxadiazole) lipids in chloroform (1 g/L) are spread on the PVA gel and vacuum dried in a desiccator for 15 minutes. At this stage, the lipids under solvent evaporation spontaneously form stacks of lipid layers supported by the dried PVA gel film. This lipid system obtained is then hydrated with 200 µL of sucrose (50-150 mM) and allowed to grow for 2 hours while remaining sealed to avoid water evaporation that would lead to sucrose solution concentration increase. The vesicle suspension is then collected and sedimented in 1 mL of glucose solution with matching concentration to have isotonic conditions. They are stored in isotonic conditions for optimal stability and osmotically deflated when preparing the sample (see below). The slight density mismatch between the sucrose solution inside the vesicle and the sucrose/glucose solution in the outer aqueous medium allows the vesicles to sediment at the bottom of the observation cell without strongly deforming them.

\textit{Sample cell preparation.} The sample cell consists in a thin glass coverslip (0.17 mm thickness, Menzel-Gläser) on top of which a self-adhesive silicon imaging chamber (CoreWell Imaging Chamber) of 0.9 mm diameter and 1.6 mm thickness is placed. The whole forms a sealed through which can then be filled with 150 $\mu$L of a glucose solution with concentration matching the vesicle outer solution concentration. 1-5 $\mu$L of concentrated GUVs solution can then be added as well as 2 $\mu$L of particles solution. The whole is then left open for an hour to allow evaporation of the water in the glucose phase and induce the deflation of the GUVs.

\textit{Optical setup.} The microscope used here is a Nikon Eclipse TE2000 confocal microscope which was used both as a standard inverted microscope for bright field and epifluorescence visualization of the sample and as a confocal microscope. Objectives used are Nikon x40 air objective, x60 water objective and x100 oil objective. Bright field illumination light source was a LHS-H100P-1 12V100W halogen lamp. The fluorescence excitation light source is the Nikon C-HGFI Intensilight, combined with the appropriate set of filters (Semrock optical filters as exciter, emitter and dichroic) to shine blue light on the sample. Bright field as well as epifluorescence acquisitions were done using a Hamamatsu C13440 Orca-Flash 4.0 digital camera. 

\textit{Acquisition and Tracking.} Videos were acquired at frequencies ranging from 25 to 100 frames per second (fps). Tracking of the Janus Cu-SiO$_2$ particles center of mass required an appropriate thresholding of the microscope acquisitions for the tracking algorithm not to be influenced by the particle orientation changes. Fitting the raw images indeed led to additional unphysical motion of the center of mass due to the orientation variations correlated with a change in aspect of the particle. To get rid of this effect, microscopy acquisitions were thresholded using ImageJ so that the particle would appear as a plain circular dark spot and the tracking was insensitive to orientation changes. The center of mass tracking of these dark spots was then achieved using the open source software Blender \textit{v 3.0.0} (Blender, Inc.). Orientation tracking was performed using a homemade ImageJ routine (Supporting Information) consisting in inverting the image and subsequently threshold appropriately in order to track the orientation of the copper coated hemisphere whose 2D projection is close to an ellipsoid.

\begin{acknowledgments}
We  wish  to  acknowledge funding from the Ecole Doctorale Physique Chimie-Physique of Strasbourg, Agence Nationale de la Recherche EDEM (Grant No.ANR-21-CE06-0042-01)  and  ITI  HiFunMat  (Université  de Strasbourg).
\end{acknowledgments}

\bibliography{apssamp}% Produces the bibliography via BibTeX.

\begin{thebibliography}{50}%
\makeatletter
\providecommand \@ifxundefined [1]{%
 \@ifx{#1\undefined}
}%
\providecommand \@ifnum [1]{%
 \ifnum #1\expandafter \@firstoftwo
 \else \expandafter \@secondoftwo
 \fi
}%
\providecommand \@ifx [1]{%
 \ifx #1\expandafter \@firstoftwo
 \else \expandafter \@secondoftwo
 \fi
}%
\providecommand \natexlab [1]{#1}%
\providecommand \enquote  [1]{``#1''}%
\providecommand \bibnamefont  [1]{#1}%
\providecommand \bibfnamefont [1]{#1}%
\providecommand \citenamefont [1]{#1}%
\providecommand \href@noop [0]{\@secondoftwo}%
\providecommand \href [0]{\begingroup \@sanitize@url \@href}%
\providecommand \@href[1]{\@@startlink{#1}\@@href}%
\providecommand \@@href[1]{\endgroup#1\@@endlink}%
\providecommand \@sanitize@url [0]{\catcode `\\12\catcode `\$12\catcode `\&12\catcode `\#12\catcode `\^12\catcode `\_12\catcode `\%12\relax}%
\providecommand \@@startlink[1]{}%
\providecommand \@@endlink[0]{}%
\providecommand \url  [0]{\begingroup\@sanitize@url \@url }%
\providecommand \@url [1]{\endgroup\@href {#1}{\urlprefix }}%
\providecommand \urlprefix  [0]{URL }%
\providecommand \Eprint [0]{\href }%
\providecommand \doibase [0]{https://doi.org/}%
\providecommand \selectlanguage [0]{\@gobble}%
\providecommand \bibinfo  [0]{\@secondoftwo}%
\providecommand \bibfield  [0]{\@secondoftwo}%
\providecommand \translation [1]{[#1]}%
\providecommand \BibitemOpen [0]{}%
\providecommand \bibitemStop [0]{}%
\providecommand \bibitemNoStop [0]{.\EOS\space}%
\providecommand \EOS [0]{\spacefactor3000\relax}%
\providecommand \BibitemShut  [1]{\csname bibitem#1\endcsname}%
\let\auto@bib@innerbib\@empty
%</preamble>
\bibitem [{\citenamefont {Howse}\ \emph {et~al.}(2007)\citenamefont {Howse}, \citenamefont {Jones}, \citenamefont {Ryan}, \citenamefont {Gough}, \citenamefont {Vafabakhsh},\ and\ \citenamefont {Golestanian}}]{Howse2007_H2O2}%
  \BibitemOpen
  \bibfield  {author} {\bibinfo {author} {\bibfnamefont {J.~R.}\ \bibnamefont {Howse}}, \bibinfo {author} {\bibfnamefont {R.~A.~L.}\ \bibnamefont {Jones}}, \bibinfo {author} {\bibfnamefont {A.~J.}\ \bibnamefont {Ryan}}, \bibinfo {author} {\bibfnamefont {T.}~\bibnamefont {Gough}}, \bibinfo {author} {\bibfnamefont {R.}~\bibnamefont {Vafabakhsh}},\ and\ \bibinfo {author} {\bibfnamefont {R.}~\bibnamefont {Golestanian}},\ }\bibfield  {title} {\bibinfo {title} {Self-motile colloidal particles: From directed propulsion to random walk},\ }\href {https://doi.org/10.1103/PhysRevLett.99.048102} {\bibfield  {journal} {\bibinfo  {journal} {Phys. Rev. Lett.}\ }\textbf {\bibinfo {volume} {99}},\ \bibinfo {pages} {048102} (\bibinfo {year} {2007})}\BibitemShut {NoStop}%
\bibitem [{\citenamefont {Gao}\ \emph {et~al.}(2014)\citenamefont {Gao}, \citenamefont {Pei}, \citenamefont {Dong},\ and\ \citenamefont {Wang}}]{Gao2014_hydrazine}%
  \BibitemOpen
  \bibfield  {author} {\bibinfo {author} {\bibfnamefont {W.}~\bibnamefont {Gao}}, \bibinfo {author} {\bibfnamefont {A.}~\bibnamefont {Pei}}, \bibinfo {author} {\bibfnamefont {R.}~\bibnamefont {Dong}},\ and\ \bibinfo {author} {\bibfnamefont {J.}~\bibnamefont {Wang}},\ }\bibfield  {title} {\bibinfo {title} {Catalytic iridium-based janus micromotors powered by ultralow levels of chemical fuels},\ }\href {https://doi.org/10.1021/ja413002e} {\bibfield  {journal} {\bibinfo  {journal} {Journal of the American Chemical Society}\ }\textbf {\bibinfo {volume} {136}},\ \bibinfo {pages} {2276} (\bibinfo {year} {2014})},\ \bibinfo {note} {pMID: 24475997}\BibitemShut {NoStop}%
\bibitem [{\citenamefont {Feuerstein}\ \emph {et~al.}(2021)\citenamefont {Feuerstein}, \citenamefont {Biermann}, \citenamefont {Xiao}, \citenamefont {Holm},\ and\ \citenamefont {Simmchen}}]{Feuerstein2021_galvanic}%
  \BibitemOpen
  \bibfield  {author} {\bibinfo {author} {\bibfnamefont {L.}~\bibnamefont {Feuerstein}}, \bibinfo {author} {\bibfnamefont {C.~G.}\ \bibnamefont {Biermann}}, \bibinfo {author} {\bibfnamefont {Z.}~\bibnamefont {Xiao}}, \bibinfo {author} {\bibfnamefont {C.}~\bibnamefont {Holm}},\ and\ \bibinfo {author} {\bibfnamefont {J.}~\bibnamefont {Simmchen}},\ }\bibfield  {title} {\bibinfo {title} {Highly efficient active colloids driven by galvanic exchange reactions},\ }\href {https://doi.org/10.1021/jacs.1c06400} {\bibfield  {journal} {\bibinfo  {journal} {Journal of the American Chemical Society}\ }\textbf {\bibinfo {volume} {143}},\ \bibinfo {pages} {17015} (\bibinfo {year} {2021})}\BibitemShut {NoStop}%
\bibitem [{\citenamefont {Wittmann}\ \emph {et~al.}(2023)\citenamefont {Wittmann}, \citenamefont {Voigtmann},\ and\ \citenamefont {Simmchen}}]{Wittmann2023_photodeposition}%
  \BibitemOpen
  \bibfield  {author} {\bibinfo {author} {\bibfnamefont {M.}~\bibnamefont {Wittmann}}, \bibinfo {author} {\bibfnamefont {M.}~\bibnamefont {Voigtmann}},\ and\ \bibinfo {author} {\bibfnamefont {J.}~\bibnamefont {Simmchen}},\ }\bibfield  {title} {\bibinfo {title} {Active bivo4 swimmers propelled by depletion gradients caused by photodeposition},\ }\href {https://doi.org/https://doi.org/10.1002/smll.202206885} {\bibfield  {journal} {\bibinfo  {journal} {Small}\ }\textbf {\bibinfo {volume} {19}},\ \bibinfo {pages} {2206885} (\bibinfo {year} {2023})}\BibitemShut {NoStop}%
\bibitem [{\citenamefont {Wu}\ \emph {et~al.}(2017)\citenamefont {Wu}, \citenamefont {Dong}, \citenamefont {Zhang},\ and\ \citenamefont {Ren}}]{Wu2017_degradation}%
  \BibitemOpen
  \bibfield  {author} {\bibinfo {author} {\bibfnamefont {Y.}~\bibnamefont {Wu}}, \bibinfo {author} {\bibfnamefont {R.}~\bibnamefont {Dong}}, \bibinfo {author} {\bibfnamefont {Q.}~\bibnamefont {Zhang}},\ and\ \bibinfo {author} {\bibfnamefont {B.}~\bibnamefont {Ren}},\ }\bibfield  {title} {\bibinfo {title} {Dye-enhanced self-electrophoretic propulsion of light-driven tio2--au janus micromotors},\ }\href {https://doi.org/10.1007/s40820-017-0133-9} {\bibfield  {journal} {\bibinfo  {journal} {Nano-Micro Letters}\ }\textbf {\bibinfo {volume} {9}},\ \bibinfo {pages} {30} (\bibinfo {year} {2017})}\BibitemShut {NoStop}%
\bibitem [{\citenamefont {Zhang}\ \emph {et~al.}(2017)\citenamefont {Zhang}, \citenamefont {Dong}, \citenamefont {Wu}, \citenamefont {Gao}, \citenamefont {He},\ and\ \citenamefont {Ren}}]{Zhang2017_degradation}%
  \BibitemOpen
  \bibfield  {author} {\bibinfo {author} {\bibfnamefont {Q.}~\bibnamefont {Zhang}}, \bibinfo {author} {\bibfnamefont {R.}~\bibnamefont {Dong}}, \bibinfo {author} {\bibfnamefont {Y.}~\bibnamefont {Wu}}, \bibinfo {author} {\bibfnamefont {W.}~\bibnamefont {Gao}}, \bibinfo {author} {\bibfnamefont {Z.}~\bibnamefont {He}},\ and\ \bibinfo {author} {\bibfnamefont {B.}~\bibnamefont {Ren}},\ }\bibfield  {title} {\bibinfo {title} {Light-driven au-wo3@c janus micromotors for rapid photodegradation of dye pollutants},\ }\href {https://doi.org/10.1021/acsami.6b12081} {\bibfield  {journal} {\bibinfo  {journal} {ACS Applied Materials \& Interfaces}\ }\textbf {\bibinfo {volume} {9}},\ \bibinfo {pages} {4674} (\bibinfo {year} {2017})},\ \bibinfo {note} {pMID: 28097861}\BibitemShut {NoStop}%
\bibitem [{\citenamefont {Wittmann}\ \emph {et~al.}(2022)\citenamefont {Wittmann}, \citenamefont {Heckel}, \citenamefont {Wurl}, \citenamefont {Xiao}, \citenamefont {Gemming}, \citenamefont {Strassner},\ and\ \citenamefont {Simmchen}}]{Wittmann2022_amine}%
  \BibitemOpen
  \bibfield  {author} {\bibinfo {author} {\bibfnamefont {M.}~\bibnamefont {Wittmann}}, \bibinfo {author} {\bibfnamefont {S.}~\bibnamefont {Heckel}}, \bibinfo {author} {\bibfnamefont {F.}~\bibnamefont {Wurl}}, \bibinfo {author} {\bibfnamefont {Z.}~\bibnamefont {Xiao}}, \bibinfo {author} {\bibfnamefont {T.}~\bibnamefont {Gemming}}, \bibinfo {author} {\bibfnamefont {T.}~\bibnamefont {Strassner}},\ and\ \bibinfo {author} {\bibfnamefont {J.}~\bibnamefont {Simmchen}},\ }\bibfield  {title} {\bibinfo {title} {Microswimming by oxidation of dibenzylamine},\ }\href {https://doi.org/10.1039/D1CC06976D} {\bibfield  {journal} {\bibinfo  {journal} {Chem. Commun.}\ }\textbf {\bibinfo {volume} {58}},\ \bibinfo {pages} {4052} (\bibinfo {year} {2022})}\BibitemShut {NoStop}%
\bibitem [{\citenamefont {Wang}\ \emph {et~al.}(2019)\citenamefont {Wang}, \citenamefont {Dong}, \citenamefont {Wang}, \citenamefont {Xu}, \citenamefont {Chen}, \citenamefont {Liang}, \citenamefont {Ren}, \citenamefont {Gao},\ and\ \citenamefont {Cai}}]{Wang2019_glucose}%
  \BibitemOpen
  \bibfield  {author} {\bibinfo {author} {\bibfnamefont {Q.}~\bibnamefont {Wang}}, \bibinfo {author} {\bibfnamefont {R.}~\bibnamefont {Dong}}, \bibinfo {author} {\bibfnamefont {C.}~\bibnamefont {Wang}}, \bibinfo {author} {\bibfnamefont {S.}~\bibnamefont {Xu}}, \bibinfo {author} {\bibfnamefont {D.}~\bibnamefont {Chen}}, \bibinfo {author} {\bibfnamefont {Y.}~\bibnamefont {Liang}}, \bibinfo {author} {\bibfnamefont {B.}~\bibnamefont {Ren}}, \bibinfo {author} {\bibfnamefont {W.}~\bibnamefont {Gao}},\ and\ \bibinfo {author} {\bibfnamefont {Y.}~\bibnamefont {Cai}},\ }\bibfield  {title} {\bibinfo {title} {Glucose-fueled micromotors with highly efficient visible-light photocatalytic propulsion},\ }\href {https://doi.org/10.1021/acsami.8b17563} {\bibfield  {journal} {\bibinfo  {journal} {ACS Applied Materials \& Interfaces}\ }\textbf {\bibinfo {volume} {11}},\ \bibinfo {pages} {6201} (\bibinfo {year} {2019})}\BibitemShut {NoStop}%
\bibitem [{\citenamefont {Vutukuri}\ \emph {et~al.}(2020)\citenamefont {Vutukuri}, \citenamefont {Hoore}, \citenamefont {Abaurrea-Velasco}, \citenamefont {van Buren}, \citenamefont {Dutto}, \citenamefont {Auth}, \citenamefont {Fedosov}, \citenamefont {Gompper},\ and\ \citenamefont {Vermant}}]{vutukuri2020active}%
  \BibitemOpen
  \bibfield  {author} {\bibinfo {author} {\bibfnamefont {H.~R.}\ \bibnamefont {Vutukuri}}, \bibinfo {author} {\bibfnamefont {M.}~\bibnamefont {Hoore}}, \bibinfo {author} {\bibfnamefont {C.}~\bibnamefont {Abaurrea-Velasco}}, \bibinfo {author} {\bibfnamefont {L.}~\bibnamefont {van Buren}}, \bibinfo {author} {\bibfnamefont {A.}~\bibnamefont {Dutto}}, \bibinfo {author} {\bibfnamefont {T.}~\bibnamefont {Auth}}, \bibinfo {author} {\bibfnamefont {D.~A.}\ \bibnamefont {Fedosov}}, \bibinfo {author} {\bibfnamefont {G.}~\bibnamefont {Gompper}},\ and\ \bibinfo {author} {\bibfnamefont {J.}~\bibnamefont {Vermant}},\ }\bibfield  {title} {\bibinfo {title} {Active particles induce large shape deformations in giant lipid vesicles},\ }\href@noop {} {\bibfield  {journal} {\bibinfo  {journal} {Nature}\ }\textbf {\bibinfo {volume} {586}},\ \bibinfo {pages} {52} (\bibinfo {year} {2020})}\BibitemShut {NoStop}%
\bibitem [{\citenamefont {Le~Nagard}\ \emph {et~al.}(2022)\citenamefont {Le~Nagard}, \citenamefont {Brown}, \citenamefont {Dawson}, \citenamefont {Martinez}, \citenamefont {Poon},\ and\ \citenamefont {Staykova}}]{le2022encapsulated}%
  \BibitemOpen
  \bibfield  {author} {\bibinfo {author} {\bibfnamefont {L.}~\bibnamefont {Le~Nagard}}, \bibinfo {author} {\bibfnamefont {A.~T.}\ \bibnamefont {Brown}}, \bibinfo {author} {\bibfnamefont {A.}~\bibnamefont {Dawson}}, \bibinfo {author} {\bibfnamefont {V.~A.}\ \bibnamefont {Martinez}}, \bibinfo {author} {\bibfnamefont {W.~C.}\ \bibnamefont {Poon}},\ and\ \bibinfo {author} {\bibfnamefont {M.}~\bibnamefont {Staykova}},\ }\bibfield  {title} {\bibinfo {title} {Encapsulated bacteria deform lipid vesicles into flagellated swimmers},\ }\href@noop {} {\bibfield  {journal} {\bibinfo  {journal} {Proceedings of the National Academy of Sciences}\ }\textbf {\bibinfo {volume} {119}},\ \bibinfo {pages} {e2206096119} (\bibinfo {year} {2022})}\BibitemShut {NoStop}%
\bibitem [{\citenamefont {Lee}\ \emph {et~al.}(2023)\citenamefont {Lee}, \citenamefont {Sch{\"o}nh{\"o}fer},\ and\ \citenamefont {Glotzer}}]{lee2023complex}%
  \BibitemOpen
  \bibfield  {author} {\bibinfo {author} {\bibfnamefont {S.~Y.}\ \bibnamefont {Lee}}, \bibinfo {author} {\bibfnamefont {P.~W.}\ \bibnamefont {Sch{\"o}nh{\"o}fer}},\ and\ \bibinfo {author} {\bibfnamefont {S.~C.}\ \bibnamefont {Glotzer}},\ }\bibfield  {title} {\bibinfo {title} {Complex motion of steerable vesicular robots filled with active colloidal rods},\ }\href@noop {} {\bibfield  {journal} {\bibinfo  {journal} {Scientific Reports}\ }\textbf {\bibinfo {volume} {13}},\ \bibinfo {pages} {22773} (\bibinfo {year} {2023})}\BibitemShut {NoStop}%
\bibitem [{\citenamefont {Iyer}\ \emph {et~al.}(2022)\citenamefont {Iyer}, \citenamefont {Gompper},\ and\ \citenamefont {Fedosov}}]{iyer2022non}%
  \BibitemOpen
  \bibfield  {author} {\bibinfo {author} {\bibfnamefont {P.}~\bibnamefont {Iyer}}, \bibinfo {author} {\bibfnamefont {G.}~\bibnamefont {Gompper}},\ and\ \bibinfo {author} {\bibfnamefont {D.~A.}\ \bibnamefont {Fedosov}},\ }\bibfield  {title} {\bibinfo {title} {Non-equilibrium shapes and dynamics of active vesicles},\ }\href@noop {} {\bibfield  {journal} {\bibinfo  {journal} {Soft matter}\ }\textbf {\bibinfo {volume} {18}},\ \bibinfo {pages} {6868} (\bibinfo {year} {2022})}\BibitemShut {NoStop}%
\bibitem [{\citenamefont {Makarchuk}\ \emph {et~al.}(2019)\citenamefont {Makarchuk}, \citenamefont {Braz}, \citenamefont {Ara{\'u}jo}, \citenamefont {Ciric},\ and\ \citenamefont {Volpe}}]{makarchuk2019enhanced}%
  \BibitemOpen
  \bibfield  {author} {\bibinfo {author} {\bibfnamefont {S.}~\bibnamefont {Makarchuk}}, \bibinfo {author} {\bibfnamefont {V.~C.}\ \bibnamefont {Braz}}, \bibinfo {author} {\bibfnamefont {N.~A.}\ \bibnamefont {Ara{\'u}jo}}, \bibinfo {author} {\bibfnamefont {L.}~\bibnamefont {Ciric}},\ and\ \bibinfo {author} {\bibfnamefont {G.}~\bibnamefont {Volpe}},\ }\bibfield  {title} {\bibinfo {title} {Enhanced propagation of motile bacteria on surfaces due to forward scattering},\ }\href@noop {} {\bibfield  {journal} {\bibinfo  {journal} {Nature Communications}\ }\textbf {\bibinfo {volume} {10}},\ \bibinfo {pages} {4110} (\bibinfo {year} {2019})}\BibitemShut {NoStop}%
\bibitem [{\citenamefont {Simmchen}\ \emph {et~al.}(2016)\citenamefont {Simmchen}, \citenamefont {Katuri}, \citenamefont {Uspal}, \citenamefont {Popescu}, \citenamefont {Tasinkevych},\ and\ \citenamefont {Sánchez}}]{Simmchen2016}%
  \BibitemOpen
  \bibfield  {author} {\bibinfo {author} {\bibfnamefont {J.}~\bibnamefont {Simmchen}}, \bibinfo {author} {\bibfnamefont {J.}~\bibnamefont {Katuri}}, \bibinfo {author} {\bibfnamefont {W.~E.}\ \bibnamefont {Uspal}}, \bibinfo {author} {\bibfnamefont {M.~N.}\ \bibnamefont {Popescu}}, \bibinfo {author} {\bibfnamefont {M.}~\bibnamefont {Tasinkevych}},\ and\ \bibinfo {author} {\bibfnamefont {S.}~\bibnamefont {Sánchez}},\ }\bibfield  {title} {\bibinfo {title} {Topographical pathways guide chemical microswimmers},\ }\bibfield  {journal} {\bibinfo  {journal} {Nature Communications}\ }\textbf {\bibinfo {volume} {7}},\ \href {https://doi.org/10.1038/ncomms10598} {10.1038/ncomms10598} (\bibinfo {year} {2016})\BibitemShut {NoStop}%
\bibitem [{\citenamefont {van Baalen}\ \emph {et~al.}(2023)\citenamefont {van Baalen}, \citenamefont {Uspal}, \citenamefont {Popescu},\ and\ \citenamefont {Isa}}]{van2023transition}%
  \BibitemOpen
  \bibfield  {author} {\bibinfo {author} {\bibfnamefont {C.}~\bibnamefont {van Baalen}}, \bibinfo {author} {\bibfnamefont {W.~E.}\ \bibnamefont {Uspal}}, \bibinfo {author} {\bibfnamefont {M.~N.}\ \bibnamefont {Popescu}},\ and\ \bibinfo {author} {\bibfnamefont {L.}~\bibnamefont {Isa}},\ }\bibfield  {title} {\bibinfo {title} {Transition from scattering to orbiting upon increasing the fuel concentration for an active janus colloid moving at an obstacle--decorated interface},\ }\href@noop {} {\bibfield  {journal} {\bibinfo  {journal} {Soft Matter}\ }\textbf {\bibinfo {volume} {19}},\ \bibinfo {pages} {8790} (\bibinfo {year} {2023})}\BibitemShut {NoStop}%
\bibitem [{\citenamefont {Sharma}\ \emph {et~al.}(2021)\citenamefont {Sharma}, \citenamefont {Azar}, \citenamefont {Schroder}, \citenamefont {Marques},\ and\ \citenamefont {Stocco}}]{Sharma2021}%
  \BibitemOpen
  \bibfield  {author} {\bibinfo {author} {\bibfnamefont {V.}~\bibnamefont {Sharma}}, \bibinfo {author} {\bibfnamefont {E.}~\bibnamefont {Azar}}, \bibinfo {author} {\bibfnamefont {A.~P.}\ \bibnamefont {Schroder}}, \bibinfo {author} {\bibfnamefont {C.~M.}\ \bibnamefont {Marques}},\ and\ \bibinfo {author} {\bibfnamefont {A.}~\bibnamefont {Stocco}},\ }\bibfield  {title} {\bibinfo {title} {Active colloids orbiting giant vesicles},\ }\href {https://doi.org/10.1039/d0sm02183k} {\bibfield  {journal} {\bibinfo  {journal} {Soft Matter}\ }\textbf {\bibinfo {volume} {17}},\ \bibinfo {pages} {4275} (\bibinfo {year} {2021})}\BibitemShut {NoStop}%
\bibitem [{\citenamefont {Sharma}\ \emph {et~al.}(2022)\citenamefont {Sharma}, \citenamefont {Marques},\ and\ \citenamefont {Stocco}}]{Sharma2022}%
  \BibitemOpen
  \bibfield  {author} {\bibinfo {author} {\bibfnamefont {V.}~\bibnamefont {Sharma}}, \bibinfo {author} {\bibfnamefont {C.~M.}\ \bibnamefont {Marques}},\ and\ \bibinfo {author} {\bibfnamefont {A.}~\bibnamefont {Stocco}},\ }\bibfield  {title} {\bibinfo {title} {Driven engulfment of janus particles by giant vesicles in and out of thermal equilibrium},\ }\bibfield  {journal} {\bibinfo  {journal} {Nanomaterials}\ }\textbf {\bibinfo {volume} {12}},\ \href {https://doi.org/10.3390/nano12091434} {10.3390/nano12091434} (\bibinfo {year} {2022})\BibitemShut {NoStop}%
\bibitem [{\citenamefont {Campbell}\ \emph {et~al.}(2019)\citenamefont {Campbell}, \citenamefont {Ebbens}, \citenamefont {Illien},\ and\ \citenamefont {Golestanian}}]{campbell2019experimental}%
  \BibitemOpen
  \bibfield  {author} {\bibinfo {author} {\bibfnamefont {A.~I.}\ \bibnamefont {Campbell}}, \bibinfo {author} {\bibfnamefont {S.~J.}\ \bibnamefont {Ebbens}}, \bibinfo {author} {\bibfnamefont {P.}~\bibnamefont {Illien}},\ and\ \bibinfo {author} {\bibfnamefont {R.}~\bibnamefont {Golestanian}},\ }\bibfield  {title} {\bibinfo {title} {Experimental observation of flow fields around active janus spheres},\ }\href@noop {} {\bibfield  {journal} {\bibinfo  {journal} {Nature communications}\ }\textbf {\bibinfo {volume} {10}},\ \bibinfo {pages} {3952} (\bibinfo {year} {2019})}\BibitemShut {NoStop}%
\bibitem [{\citenamefont {Xiao}\ \emph {et~al.}(2022{\natexlab{a}})\citenamefont {Xiao}, \citenamefont {Ma},\ and\ \citenamefont {Wu}}]{Xiao2022}%
  \BibitemOpen
  \bibfield  {author} {\bibinfo {author} {\bibfnamefont {K.}~\bibnamefont {Xiao}}, \bibinfo {author} {\bibfnamefont {R.}~\bibnamefont {Ma}},\ and\ \bibinfo {author} {\bibfnamefont {C.-X.}\ \bibnamefont {Wu}},\ }\bibfield  {title} {\bibinfo {title} {Force-induced wrapping phase transition in activated cellular uptake},\ }\bibfield  {journal} {\bibinfo  {journal} {Physical Review E}\ }\textbf {\bibinfo {volume} {106}},\ \href {https://doi.org/10.1103/physreve.106.044411} {10.1103/physreve.106.044411} (\bibinfo {year} {2022}{\natexlab{a}})\BibitemShut {NoStop}%
\bibitem [{\citenamefont {Sharan}\ \emph {et~al.}(2022)\citenamefont {Sharan}, \citenamefont {Xiao}, \citenamefont {Mancuso}, \citenamefont {Uspal},\ and\ \citenamefont {Simmchen}}]{Sharan2022}%
  \BibitemOpen
  \bibfield  {author} {\bibinfo {author} {\bibfnamefont {P.}~\bibnamefont {Sharan}}, \bibinfo {author} {\bibfnamefont {Z.}~\bibnamefont {Xiao}}, \bibinfo {author} {\bibfnamefont {V.}~\bibnamefont {Mancuso}}, \bibinfo {author} {\bibfnamefont {W.~E.}\ \bibnamefont {Uspal}},\ and\ \bibinfo {author} {\bibfnamefont {J.}~\bibnamefont {Simmchen}},\ }\bibfield  {title} {\bibinfo {title} {Upstream rheotaxis of catalytic janus spheres},\ }\href {https://doi.org/10.1021/acsnano.1c11204} {\bibfield  {journal} {\bibinfo  {journal} {ACS Nano}\ }\textbf {\bibinfo {volume} {16}},\ \bibinfo {pages} {4599} (\bibinfo {year} {2022})}\BibitemShut {NoStop}%
\bibitem [{\citenamefont {Sharan}\ \emph {et~al.}(2023)\citenamefont {Sharan}, \citenamefont {Daddi-Moussa-Ider}, \citenamefont {Agudo-Canalejo}, \citenamefont {Golestanian},\ and\ \citenamefont {Simmchen}}]{Sharan2023}%
  \BibitemOpen
  \bibfield  {author} {\bibinfo {author} {\bibfnamefont {P.}~\bibnamefont {Sharan}}, \bibinfo {author} {\bibfnamefont {A.}~\bibnamefont {Daddi-Moussa-Ider}}, \bibinfo {author} {\bibfnamefont {J.}~\bibnamefont {Agudo-Canalejo}}, \bibinfo {author} {\bibfnamefont {R.}~\bibnamefont {Golestanian}},\ and\ \bibinfo {author} {\bibfnamefont {J.}~\bibnamefont {Simmchen}},\ }\bibfield  {title} {\bibinfo {title} {Pair interaction between two catalytically active colloids},\ }\href {https://doi.org/https://doi.org/10.1002/smll.202300817} {\bibfield  {journal} {\bibinfo  {journal} {Small}\ }\textbf {\bibinfo {volume} {19}},\ \bibinfo {pages} {2300817} (\bibinfo {year} {2023})}\BibitemShut {NoStop}%
\bibitem [{\citenamefont {Lauga}(2020)}]{Lauga_2020}%
  \BibitemOpen
  \bibfield  {author} {\bibinfo {author} {\bibfnamefont {E.}~\bibnamefont {Lauga}},\ }\href@noop {} {\emph {\bibinfo {title} {The Fluid Dynamics of Cell Motility}}},\ Cambridge Texts in Applied Mathematics\ (\bibinfo  {publisher} {Cambridge University Press},\ \bibinfo {year} {2020})\BibitemShut {NoStop}%
\bibitem [{\citenamefont {Ketzetzi}\ \emph {et~al.}(2020)\citenamefont {Ketzetzi}, \citenamefont {Graaf},\ and\ \citenamefont {Kraft}}]{Ketzetzi2020}%
  \BibitemOpen
  \bibfield  {author} {\bibinfo {author} {\bibfnamefont {S.}~\bibnamefont {Ketzetzi}}, \bibinfo {author} {\bibfnamefont {J.~D.}\ \bibnamefont {Graaf}},\ and\ \bibinfo {author} {\bibfnamefont {D.~J.}\ \bibnamefont {Kraft}},\ }\bibfield  {title} {\bibinfo {title} {Diffusion-based height analysis reveals robust microswimmer-wall separation},\ }\bibfield  {journal} {\bibinfo  {journal} {Physical Review Letters}\ }\textbf {\bibinfo {volume} {125}},\ \href {https://doi.org/10.1103/PhysRevLett.125.238001} {10.1103/PhysRevLett.125.238001} (\bibinfo {year} {2020})\BibitemShut {NoStop}%
\bibitem [{\citenamefont {Block}(1991)}]{BLOCK1997}%
  \BibitemOpen
  \bibfield  {author} {\bibinfo {author} {\bibfnamefont {E.~R.}\ \bibnamefont {Block}},\ }\bibfield  {title} {\bibinfo {title} {Hydrogen peroxide alters the physical state and function of the plasma membrane of pulmonary artery endothelial cells},\ }\href {https://doi.org/https://doi.org/10.1002/jcp.1041460305} {\bibfield  {journal} {\bibinfo  {journal} {Journal of Cellular Physiology}\ }\textbf {\bibinfo {volume} {146}},\ \bibinfo {pages} {362} (\bibinfo {year} {1991})}\BibitemShut {NoStop}%
\bibitem [{\citenamefont {Valen}\ \emph {et~al.}(1999)\citenamefont {Valen}, \citenamefont {Sondén}, \citenamefont {Vaage}, \citenamefont {Malm},\ and\ \citenamefont {Kjellström}}]{VALEN19991480}%
  \BibitemOpen
  \bibfield  {author} {\bibinfo {author} {\bibfnamefont {G.}~\bibnamefont {Valen}}, \bibinfo {author} {\bibfnamefont {A.}~\bibnamefont {Sondén}}, \bibinfo {author} {\bibfnamefont {J.}~\bibnamefont {Vaage}}, \bibinfo {author} {\bibfnamefont {E.}~\bibnamefont {Malm}},\ and\ \bibinfo {author} {\bibfnamefont {B.}~\bibnamefont {Kjellström}},\ }\bibfield  {title} {\bibinfo {title} {Hydrogen peroxide induces endothelial cell atypia and cytoskeleton depolymerization},\ }\href {https://doi.org/https://doi.org/10.1016/S0891-5849(99)00009-X} {\bibfield  {journal} {\bibinfo  {journal} {Free Radical Biology and Medicine}\ }\textbf {\bibinfo {volume} {26}},\ \bibinfo {pages} {1480} (\bibinfo {year} {1999})}\BibitemShut {NoStop}%
\bibitem [{\citenamefont {Dai}\ \emph {et~al.}(1997)\citenamefont {Dai}, \citenamefont {Ting-Beall},\ and\ \citenamefont {Sheetz}}]{Dai1997}%
  \BibitemOpen
  \bibfield  {author} {\bibinfo {author} {\bibfnamefont {J.}~\bibnamefont {Dai}}, \bibinfo {author} {\bibfnamefont {H.~P.}\ \bibnamefont {Ting-Beall}},\ and\ \bibinfo {author} {\bibfnamefont {M.~P.}\ \bibnamefont {Sheetz}},\ }\bibfield  {title} {\bibinfo {title} {The secretion-coupled endocytosis correlates with membrane tension changes in rbl 2h3 cells},\ }\href {http://rupress.org/jgp/article-pdf/110/1/1/1770270/gp-7450.pdf} {\bibfield  {journal} {\bibinfo  {journal} {J. Gen. Physiol}\ }\textbf {\bibinfo {volume} {110}} (\bibinfo {year} {1997})}\BibitemShut {NoStop}%
\bibitem [{\citenamefont {Pontes}\ \emph {et~al.}(2017)\citenamefont {Pontes}, \citenamefont {Monzo},\ and\ \citenamefont {Gauthier}}]{PONTES201730}%
  \BibitemOpen
  \bibfield  {author} {\bibinfo {author} {\bibfnamefont {B.}~\bibnamefont {Pontes}}, \bibinfo {author} {\bibfnamefont {P.}~\bibnamefont {Monzo}},\ and\ \bibinfo {author} {\bibfnamefont {N.~C.}\ \bibnamefont {Gauthier}},\ }\bibfield  {title} {\bibinfo {title} {Membrane tension: A challenging but universal physical parameter in cell biology},\ }\href {https://doi.org/https://doi.org/10.1016/j.semcdb.2017.08.030} {\bibfield  {journal} {\bibinfo  {journal} {Seminars in Cell and Developmental Biology}\ }\textbf {\bibinfo {volume} {71}},\ \bibinfo {pages} {30} (\bibinfo {year} {2017})},\ \bibinfo {note} {mechanosensing: from molecules to tissues}\BibitemShut {NoStop}%
\bibitem [{\citenamefont {Meinel}\ \emph {et~al.}(2014)\citenamefont {Meinel}, \citenamefont {Tränkle}, \citenamefont {Römer},\ and\ \citenamefont {Rohrbach}}]{Meinel2014}%
  \BibitemOpen
  \bibfield  {author} {\bibinfo {author} {\bibfnamefont {A.}~\bibnamefont {Meinel}}, \bibinfo {author} {\bibfnamefont {B.}~\bibnamefont {Tränkle}}, \bibinfo {author} {\bibfnamefont {W.}~\bibnamefont {Römer}},\ and\ \bibinfo {author} {\bibfnamefont {A.}~\bibnamefont {Rohrbach}},\ }\bibfield  {title} {\bibinfo {title} {Induced phagocytic particle uptake into a giant unilamellar vesicle},\ }\href {https://doi.org/10.1039/c3sm52964a} {\bibfield  {journal} {\bibinfo  {journal} {Soft Matter}\ }\textbf {\bibinfo {volume} {10}},\ \bibinfo {pages} {3667} (\bibinfo {year} {2014})}\BibitemShut {NoStop}%
\bibitem [{\citenamefont {Spanke}\ \emph {et~al.}(2020)\citenamefont {Spanke}, \citenamefont {Style}, \citenamefont {François-Martin}, \citenamefont {Feofilova}, \citenamefont {Eisentraut}, \citenamefont {Kress}, \citenamefont {Agudo-Canalejo},\ and\ \citenamefont {Dufresne}}]{Spanke2020}%
  \BibitemOpen
  \bibfield  {author} {\bibinfo {author} {\bibfnamefont {H.~T.}\ \bibnamefont {Spanke}}, \bibinfo {author} {\bibfnamefont {R.~W.}\ \bibnamefont {Style}}, \bibinfo {author} {\bibfnamefont {C.}~\bibnamefont {François-Martin}}, \bibinfo {author} {\bibfnamefont {M.}~\bibnamefont {Feofilova}}, \bibinfo {author} {\bibfnamefont {M.}~\bibnamefont {Eisentraut}}, \bibinfo {author} {\bibfnamefont {H.}~\bibnamefont {Kress}}, \bibinfo {author} {\bibfnamefont {J.}~\bibnamefont {Agudo-Canalejo}},\ and\ \bibinfo {author} {\bibfnamefont {E.~R.}\ \bibnamefont {Dufresne}},\ }\bibfield  {title} {\bibinfo {title} {Wrapping of microparticles by floppy lipid vesicles},\ }\bibfield  {journal} {\bibinfo  {journal} {Physical Review Letters}\ }\textbf {\bibinfo {volume} {125}},\ \href {https://doi.org/10.1103/PhysRevLett.125.198102} {10.1103/PhysRevLett.125.198102} (\bibinfo {year} {2020})\BibitemShut {NoStop}%
\bibitem [{\citenamefont {Fessler}\ \emph {et~al.}(2023)\citenamefont {Fessler}, \citenamefont {Sharma}, \citenamefont {Muller},\ and\ \citenamefont {Stocco}}]{Fessler2023}%
  \BibitemOpen
  \bibfield  {author} {\bibinfo {author} {\bibfnamefont {F.}~\bibnamefont {Fessler}}, \bibinfo {author} {\bibfnamefont {V.}~\bibnamefont {Sharma}}, \bibinfo {author} {\bibfnamefont {P.}~\bibnamefont {Muller}},\ and\ \bibinfo {author} {\bibfnamefont {A.}~\bibnamefont {Stocco}},\ }\bibfield  {title} {\bibinfo {title} {Entry of microparticles into giant lipid vesicles by optical tweezers},\ }\href {https://doi.org/10.1103/PhysRevE.107.L052601} {\bibfield  {journal} {\bibinfo  {journal} {Phys. Rev. E}\ }\textbf {\bibinfo {volume} {107}},\ \bibinfo {pages} {L052601} (\bibinfo {year} {2023})}\BibitemShut {NoStop}%
\bibitem [{\citenamefont {Das}\ \emph {et~al.}(2015)\citenamefont {Das}, \citenamefont {Garg}, \citenamefont {Campbell}, \citenamefont {Howse}, \citenamefont {Sen}, \citenamefont {Velegol}, \citenamefont {Golestanian},\ and\ \citenamefont {Ebbens}}]{Das2015}%
  \BibitemOpen
  \bibfield  {author} {\bibinfo {author} {\bibfnamefont {S.}~\bibnamefont {Das}}, \bibinfo {author} {\bibfnamefont {A.}~\bibnamefont {Garg}}, \bibinfo {author} {\bibfnamefont {A.~I.}\ \bibnamefont {Campbell}}, \bibinfo {author} {\bibfnamefont {J.}~\bibnamefont {Howse}}, \bibinfo {author} {\bibfnamefont {A.}~\bibnamefont {Sen}}, \bibinfo {author} {\bibfnamefont {D.}~\bibnamefont {Velegol}}, \bibinfo {author} {\bibfnamefont {R.}~\bibnamefont {Golestanian}},\ and\ \bibinfo {author} {\bibfnamefont {S.~J.}\ \bibnamefont {Ebbens}},\ }\bibfield  {title} {\bibinfo {title} {Boundaries can steer active janus spheres},\ }\bibfield  {journal} {\bibinfo  {journal} {Nature Communications}\ }\textbf {\bibinfo {volume} {6}},\ \href {https://doi.org/10.1038/ncomms9999} {10.1038/ncomms9999} (\bibinfo {year} {2015})\BibitemShut {NoStop}%
\bibitem [{\citenamefont {Spagnolie}\ and\ \citenamefont {Lauga}(2012)}]{Spagnolie2012}%
  \BibitemOpen
  \bibfield  {author} {\bibinfo {author} {\bibfnamefont {S.~E.}\ \bibnamefont {Spagnolie}}\ and\ \bibinfo {author} {\bibfnamefont {E.}~\bibnamefont {Lauga}},\ }\bibfield  {title} {\bibinfo {title} {Hydrodynamics of self-propulsion near a boundary: predictions and accuracy of far-field approximations},\ }\href {https://doi.org/10.1017/jfm.2012.101} {\bibfield  {journal} {\bibinfo  {journal} {Journal of Fluid Mechanics}\ }\textbf {\bibinfo {volume} {700}},\ \bibinfo {pages} {105} (\bibinfo {year} {2012})}\BibitemShut {NoStop}%
\bibitem [{\citenamefont {Spagnolie}\ \emph {et~al.}(2015)\citenamefont {Spagnolie}, \citenamefont {Moreno-Flores}, \citenamefont {Bartolo},\ and\ \citenamefont {Lauga}}]{Spagnolie2015}%
  \BibitemOpen
  \bibfield  {author} {\bibinfo {author} {\bibfnamefont {S.~E.}\ \bibnamefont {Spagnolie}}, \bibinfo {author} {\bibfnamefont {G.~R.}\ \bibnamefont {Moreno-Flores}}, \bibinfo {author} {\bibfnamefont {D.}~\bibnamefont {Bartolo}},\ and\ \bibinfo {author} {\bibfnamefont {E.}~\bibnamefont {Lauga}},\ }\bibfield  {title} {\bibinfo {title} {Geometric capture and escape of a microswimmer colliding with an obstacle},\ }\href {https://doi.org/10.1039/c4sm02785j} {\bibfield  {journal} {\bibinfo  {journal} {Soft Matter}\ }\textbf {\bibinfo {volume} {11}},\ \bibinfo {pages} {3396} (\bibinfo {year} {2015})}\BibitemShut {NoStop}%
\bibitem [{\citenamefont {Radier}\ \emph {et~al.}(1995)\citenamefont {Radier}, \citenamefont {Feder}, \citenamefont {Strey},\ and\ \citenamefont {Sackmann}}]{Radier1995}%
  \BibitemOpen
  \bibfield  {author} {\bibinfo {author} {\bibfnamefont {J.~O.}\ \bibnamefont {Radier}}, \bibinfo {author} {\bibfnamefont {T.~J.}\ \bibnamefont {Feder}}, \bibinfo {author} {\bibfnamefont {H.~H.}\ \bibnamefont {Strey}},\ and\ \bibinfo {author} {\bibfnamefont {E.}~\bibnamefont {Sackmann}},\ }\bibfield  {title} {\bibinfo {title} {Fluctuation analysis of tension-controlled undulation forces between giant vesicles and solid substrates},\ }\href@noop {} {\bibfield  {journal} {\bibinfo  {journal} {Phys. Rev. E}\ }\textbf {\bibinfo {volume} {51}} (\bibinfo {year} {1995})}\BibitemShut {NoStop}%
\bibitem [{\citenamefont {Cantat}\ \emph {et~al.}(2003)\citenamefont {Cantat}, \citenamefont {Kassner},\ and\ \citenamefont {Misbah}}]{Cantat2003}%
  \BibitemOpen
  \bibfield  {author} {\bibinfo {author} {\bibfnamefont {I.}~\bibnamefont {Cantat}}, \bibinfo {author} {\bibfnamefont {K.}~\bibnamefont {Kassner}},\ and\ \bibinfo {author} {\bibfnamefont {C.}~\bibnamefont {Misbah}},\ }\bibfield  {title} {\bibinfo {title} {Vesicles in haptotaxis with hydrodynamical dissipation},\ }\href {https://doi.org/10.1140/epje/e2003-00022-1} {\bibfield  {journal} {\bibinfo  {journal} {European Physical Journal E}\ }\textbf {\bibinfo {volume} {10}},\ \bibinfo {pages} {175} (\bibinfo {year} {2003})}\BibitemShut {NoStop}%
\bibitem [{\citenamefont {Bernard}\ \emph {et~al.}(2000)\citenamefont {Bernard}, \citenamefont {Guedeau-Boudeville}, \citenamefont {Jullien},\ and\ \citenamefont {Meglio}}]{Bernard2000}%
  \BibitemOpen
  \bibfield  {author} {\bibinfo {author} {\bibfnamefont {A.~L.}\ \bibnamefont {Bernard}}, \bibinfo {author} {\bibfnamefont {M.~A.}\ \bibnamefont {Guedeau-Boudeville}}, \bibinfo {author} {\bibfnamefont {L.}~\bibnamefont {Jullien}},\ and\ \bibinfo {author} {\bibfnamefont {J.~M.~D.}\ \bibnamefont {Meglio}},\ }\bibfield  {title} {\bibinfo {title} {Strong adhesion of giant vesicles on surfaces: dynamics and permeability},\ }\href {https://doi.org/10.1021/la991341x} {\bibfield  {journal} {\bibinfo  {journal} {Langmuir}\ }\textbf {\bibinfo {volume} {16}},\ \bibinfo {pages} {6809} (\bibinfo {year} {2000})}\BibitemShut {NoStop}%
\bibitem [{\citenamefont {Spanke}\ \emph {et~al.}(2022)\citenamefont {Spanke}, \citenamefont {Agudo-Canalejo}, \citenamefont {Tran}, \citenamefont {Style},\ and\ \citenamefont {Dufresne}}]{Spanke2022}%
  \BibitemOpen
  \bibfield  {author} {\bibinfo {author} {\bibfnamefont {H.~T.}\ \bibnamefont {Spanke}}, \bibinfo {author} {\bibfnamefont {J.}~\bibnamefont {Agudo-Canalejo}}, \bibinfo {author} {\bibfnamefont {D.}~\bibnamefont {Tran}}, \bibinfo {author} {\bibfnamefont {R.~W.}\ \bibnamefont {Style}},\ and\ \bibinfo {author} {\bibfnamefont {E.~R.}\ \bibnamefont {Dufresne}},\ }\bibfield  {title} {\bibinfo {title} {Dynamics of spontaneous wrapping of microparticles by floppy lipid membranes},\ }\href {https://doi.org/10.1103/PhysRevResearch.4.023080} {\bibfield  {journal} {\bibinfo  {journal} {Phys. Rev. Res.}\ }\textbf {\bibinfo {volume} {4}},\ \bibinfo {pages} {023080} (\bibinfo {year} {2022})}\BibitemShut {NoStop}%
\bibitem [{\citenamefont {Lipowsky}\ and\ \citenamefont {Seifertt}(1991)}]{Lipowsky1991}%
  \BibitemOpen
  \bibfield  {author} {\bibinfo {author} {\bibfnamefont {R.}~\bibnamefont {Lipowsky}}\ and\ \bibinfo {author} {\bibfnamefont {U.}~\bibnamefont {Seifertt}},\ }\bibfield  {title} {\bibinfo {title} {Adhesion of membranes: A theoretical perspective},\ }\href@noop {} {\bibfield  {journal} {\bibinfo  {journal} {LANGMUIR}\ ,\ \bibinfo {pages} {7}} (\bibinfo {year} {1991})}\BibitemShut {NoStop}%
\bibitem [{\citenamefont {Gruhn}\ \emph {et~al.}(2007)\citenamefont {Gruhn}, \citenamefont {Franke}, \citenamefont {Dimova},\ and\ \citenamefont {Lipowsky}}]{Gruhn2007}%
  \BibitemOpen
  \bibfield  {author} {\bibinfo {author} {\bibfnamefont {T.}~\bibnamefont {Gruhn}}, \bibinfo {author} {\bibfnamefont {T.}~\bibnamefont {Franke}}, \bibinfo {author} {\bibfnamefont {R.}~\bibnamefont {Dimova}},\ and\ \bibinfo {author} {\bibfnamefont {R.}~\bibnamefont {Lipowsky}},\ }\bibfield  {title} {\bibinfo {title} {Novel method for measuring the adhesion energy of vesicles},\ }\href {https://doi.org/10.1021/la063123r} {\bibfield  {journal} {\bibinfo  {journal} {Langmuir}\ }\textbf {\bibinfo {volume} {23}},\ \bibinfo {pages} {5423} (\bibinfo {year} {2007})}\BibitemShut {NoStop}%
\bibitem [{\citenamefont {Steinkühler}\ \emph {et~al.}(2016)\citenamefont {Steinkühler}, \citenamefont {Agudo-Canalejo}, \citenamefont {Lipowsky},\ and\ \citenamefont {Dimova}}]{STEINKUHLER20161454}%
  \BibitemOpen
  \bibfield  {author} {\bibinfo {author} {\bibfnamefont {J.}~\bibnamefont {Steinkühler}}, \bibinfo {author} {\bibfnamefont {J.}~\bibnamefont {Agudo-Canalejo}}, \bibinfo {author} {\bibfnamefont {R.}~\bibnamefont {Lipowsky}},\ and\ \bibinfo {author} {\bibfnamefont {R.}~\bibnamefont {Dimova}},\ }\bibfield  {title} {\bibinfo {title} {Modulating vesicle adhesion by electric fields},\ }\href {https://doi.org/https://doi.org/10.1016/j.bpj.2016.08.029} {\bibfield  {journal} {\bibinfo  {journal} {Biophysical Journal}\ }\textbf {\bibinfo {volume} {111}},\ \bibinfo {pages} {1454} (\bibinfo {year} {2016})}\BibitemShut {NoStop}%
\bibitem [{\citenamefont {Klasczyk}\ \emph {et~al.}(2010)\citenamefont {Klasczyk}, \citenamefont {Knecht}, \citenamefont {Lipowsky},\ and\ \citenamefont {Dimova}}]{POPCcharge}%
  \BibitemOpen
  \bibfield  {author} {\bibinfo {author} {\bibfnamefont {B.}~\bibnamefont {Klasczyk}}, \bibinfo {author} {\bibfnamefont {V.}~\bibnamefont {Knecht}}, \bibinfo {author} {\bibfnamefont {R.}~\bibnamefont {Lipowsky}},\ and\ \bibinfo {author} {\bibfnamefont {R.}~\bibnamefont {Dimova}},\ }\bibfield  {title} {\bibinfo {title} {Interactions of alkali metal chlorides with phosphatidylcholine vesicles},\ }\href {https://doi.org/10.1021/la103631y} {\bibfield  {journal} {\bibinfo  {journal} {Langmuir}\ }\textbf {\bibinfo {volume} {26}},\ \bibinfo {pages} {18951} (\bibinfo {year} {2010})},\ \bibinfo {note} {pMID: 21114263},\ \Eprint {https://arxiv.org/abs/https://doi.org/10.1021/la103631y} {https://doi.org/10.1021/la103631y} \BibitemShut {NoStop}%
\bibitem [{\citenamefont {Jiang}\ \emph {et~al.}(2018)\citenamefont {Jiang}, \citenamefont {Zhang}, \citenamefont {Zhou}, \citenamefont {Li}, \citenamefont {Hu}, \citenamefont {Zhu}, \citenamefont {Tan}, \citenamefont {Chang}, \citenamefont {Lü},\ and\ \citenamefont {Song}}]{Jiang2018}%
  \BibitemOpen
  \bibfield  {author} {\bibinfo {author} {\bibfnamefont {X.}~\bibnamefont {Jiang}}, \bibinfo {author} {\bibfnamefont {J.}~\bibnamefont {Zhang}}, \bibinfo {author} {\bibfnamefont {B.}~\bibnamefont {Zhou}}, \bibinfo {author} {\bibfnamefont {P.}~\bibnamefont {Li}}, \bibinfo {author} {\bibfnamefont {X.}~\bibnamefont {Hu}}, \bibinfo {author} {\bibfnamefont {Z.}~\bibnamefont {Zhu}}, \bibinfo {author} {\bibfnamefont {Y.}~\bibnamefont {Tan}}, \bibinfo {author} {\bibfnamefont {C.}~\bibnamefont {Chang}}, \bibinfo {author} {\bibfnamefont {J.}~\bibnamefont {Lü}},\ and\ \bibinfo {author} {\bibfnamefont {B.}~\bibnamefont {Song}},\ }\bibfield  {title} {\bibinfo {title} {Anomalous behavior of membrane fluidity caused by copper-copper bond coupled phospholipids},\ }\bibfield  {journal} {\bibinfo  {journal} {Scientific Reports}\ }\textbf {\bibinfo {volume} {8}},\ \href {https://doi.org/10.1038/s41598-018-32322-4} {10.1038/s41598-018-32322-4} (\bibinfo {year} {2018})\BibitemShut {NoStop}%
\bibitem [{\citenamefont {Poyton}\ \emph {et~al.}(2016)\citenamefont {Poyton}, \citenamefont {Sendecki}, \citenamefont {Cong},\ and\ \citenamefont {Cremer}}]{Poyton2016}%
  \BibitemOpen
  \bibfield  {author} {\bibinfo {author} {\bibfnamefont {M.~F.}\ \bibnamefont {Poyton}}, \bibinfo {author} {\bibfnamefont {A.~M.}\ \bibnamefont {Sendecki}}, \bibinfo {author} {\bibfnamefont {X.}~\bibnamefont {Cong}},\ and\ \bibinfo {author} {\bibfnamefont {P.~S.}\ \bibnamefont {Cremer}},\ }\bibfield  {title} {\bibinfo {title} {Cu2+ binds to phosphatidylethanolamine and increases oxidation in lipid membranes},\ }\href {https://doi.org/10.1021/jacs.5b11561} {\bibfield  {journal} {\bibinfo  {journal} {Journal of the American Chemical Society}\ }\textbf {\bibinfo {volume} {138}},\ \bibinfo {pages} {1584} (\bibinfo {year} {2016})}\BibitemShut {NoStop}%
\bibitem [{\citenamefont {H\"anggi}\ \emph {et~al.}(1990)\citenamefont {H\"anggi}, \citenamefont {Talkner},\ and\ \citenamefont {Borkovec}}]{Kramers}%
  \BibitemOpen
  \bibfield  {author} {\bibinfo {author} {\bibfnamefont {P.}~\bibnamefont {H\"anggi}}, \bibinfo {author} {\bibfnamefont {P.}~\bibnamefont {Talkner}},\ and\ \bibinfo {author} {\bibfnamefont {M.}~\bibnamefont {Borkovec}},\ }\bibfield  {title} {\bibinfo {title} {Reaction-rate theory: fifty years after kramers},\ }\href {https://doi.org/10.1103/RevModPhys.62.251} {\bibfield  {journal} {\bibinfo  {journal} {Rev. Mod. Phys.}\ }\textbf {\bibinfo {volume} {62}},\ \bibinfo {pages} {251} (\bibinfo {year} {1990})}\BibitemShut {NoStop}%
\bibitem [{\citenamefont {Brown}\ \emph {et~al.}(2016)\citenamefont {Brown}, \citenamefont {Vladescu}, \citenamefont {Dawson}, \citenamefont {Vissers}, \citenamefont {Schwarz-Linek}, \citenamefont {Lintuvuori},\ and\ \citenamefont {Poon}}]{Brown2016}%
  \BibitemOpen
  \bibfield  {author} {\bibinfo {author} {\bibfnamefont {A.~T.}\ \bibnamefont {Brown}}, \bibinfo {author} {\bibfnamefont {I.~D.}\ \bibnamefont {Vladescu}}, \bibinfo {author} {\bibfnamefont {A.}~\bibnamefont {Dawson}}, \bibinfo {author} {\bibfnamefont {T.}~\bibnamefont {Vissers}}, \bibinfo {author} {\bibfnamefont {J.}~\bibnamefont {Schwarz-Linek}}, \bibinfo {author} {\bibfnamefont {J.~S.}\ \bibnamefont {Lintuvuori}},\ and\ \bibinfo {author} {\bibfnamefont {W.~C.~K.}\ \bibnamefont {Poon}},\ }\bibfield  {title} {\bibinfo {title} {Swimming in a crystal †},\ }\href {https://doi.org/10.7488/DS/304} {\bibfield  {journal} {\bibinfo  {journal} {Soft Matter}\ }\textbf {\bibinfo {volume} {12}},\ \bibinfo {pages} {131} (\bibinfo {year} {2016})}\BibitemShut {NoStop}%
\bibitem [{\citenamefont {Xiao}\ \emph {et~al.}(2022{\natexlab{b}})\citenamefont {Xiao}, \citenamefont {Ma},\ and\ \citenamefont {Wu}}]{Wu2022}%
  \BibitemOpen
  \bibfield  {author} {\bibinfo {author} {\bibfnamefont {K.}~\bibnamefont {Xiao}}, \bibinfo {author} {\bibfnamefont {R.}~\bibnamefont {Ma}},\ and\ \bibinfo {author} {\bibfnamefont {C.-X.}\ \bibnamefont {Wu}},\ }\bibfield  {title} {\bibinfo {title} {Force-induced wrapping phase transition in activated cellular uptake},\ }\href {https://doi.org/10.1103/PhysRevE.106.044411} {\bibfield  {journal} {\bibinfo  {journal} {Phys. Rev. E}\ }\textbf {\bibinfo {volume} {106}},\ \bibinfo {pages} {044411} (\bibinfo {year} {2022}{\natexlab{b}})}\BibitemShut {NoStop}%
\bibitem [{\citenamefont {Deserno}(2004)}]{Deserno2004}%
  \BibitemOpen
  \bibfield  {author} {\bibinfo {author} {\bibfnamefont {M.}~\bibnamefont {Deserno}},\ }\bibfield  {title} {\bibinfo {title} {Elastic deformation of a fluid membrane upon colloid binding},\ }\href {https://doi.org/10.1103/PhysRevE.69.031903} {\bibfield  {journal} {\bibinfo  {journal} {Phys. Rev. E}\ }\textbf {\bibinfo {volume} {69}},\ \bibinfo {pages} {031903} (\bibinfo {year} {2004})}\BibitemShut {NoStop}%
\bibitem [{\citenamefont {Faizi}\ \emph {et~al.}(2022)\citenamefont {Faizi}, \citenamefont {Tsui}, \citenamefont {Dimova},\ and\ \citenamefont {Vlahovska}}]{Faizi2022}%
  \BibitemOpen
  \bibfield  {author} {\bibinfo {author} {\bibfnamefont {H.~A.}\ \bibnamefont {Faizi}}, \bibinfo {author} {\bibfnamefont {A.}~\bibnamefont {Tsui}}, \bibinfo {author} {\bibfnamefont {R.}~\bibnamefont {Dimova}},\ and\ \bibinfo {author} {\bibfnamefont {P.~M.}\ \bibnamefont {Vlahovska}},\ }\bibfield  {title} {\bibinfo {title} {Bending rigidity, capacitance, and shear viscosity of giant vesicle membranes prepared by spontaneous swelling, electroformation, gel-assisted, and phase transfer methods: A comparative study},\ }\href {https://doi.org/10.1021/acs.langmuir.2c01402} {\bibfield  {journal} {\bibinfo  {journal} {Langmuir}\ }\textbf {\bibinfo {volume} {38}},\ \bibinfo {pages} {10548} (\bibinfo {year} {2022})}\BibitemShut {NoStop}%
\bibitem [{\citenamefont {Döbereiner}\ \emph {et~al.}(1999)\citenamefont {Döbereiner}, \citenamefont {Selchow},\ and\ \citenamefont {Lipowsky}}]{Gunther1999}%
  \BibitemOpen
  \bibfield  {author} {\bibinfo {author} {\bibfnamefont {H.~G.}\ \bibnamefont {Döbereiner}}, \bibinfo {author} {\bibfnamefont {O.}~\bibnamefont {Selchow}},\ and\ \bibinfo {author} {\bibfnamefont {R.}~\bibnamefont {Lipowsky}},\ }\bibfield  {title} {\bibinfo {title} {Spontaneous curvature of fluid vesicles induced by trans-bilayer sugar asymmetry},\ }\href {https://doi.org/10.1007/s002490050197} {\bibfield  {journal} {\bibinfo  {journal} {European Biophysics Journal}\ }\textbf {\bibinfo {volume} {28}},\ \bibinfo {pages} {174} (\bibinfo {year} {1999})}\BibitemShut {NoStop}%
\bibitem [{\citenamefont {Weinberger}\ \emph {et~al.}(2013)\citenamefont {Weinberger}, \citenamefont {Tsai}, \citenamefont {Koenderink}, \citenamefont {Schmidt}, \citenamefont {Itri}, \citenamefont {Meier}, \citenamefont {Schmatko}, \citenamefont {Schröder},\ and\ \citenamefont {Marques}}]{Weinberger2013}%
  \BibitemOpen
  \bibfield  {author} {\bibinfo {author} {\bibfnamefont {A.}~\bibnamefont {Weinberger}}, \bibinfo {author} {\bibfnamefont {F.~C.}\ \bibnamefont {Tsai}}, \bibinfo {author} {\bibfnamefont {G.~H.}\ \bibnamefont {Koenderink}}, \bibinfo {author} {\bibfnamefont {T.~F.}\ \bibnamefont {Schmidt}}, \bibinfo {author} {\bibfnamefont {R.}~\bibnamefont {Itri}}, \bibinfo {author} {\bibfnamefont {W.}~\bibnamefont {Meier}}, \bibinfo {author} {\bibfnamefont {T.}~\bibnamefont {Schmatko}}, \bibinfo {author} {\bibfnamefont {A.}~\bibnamefont {Schröder}},\ and\ \bibinfo {author} {\bibfnamefont {C.}~\bibnamefont {Marques}},\ }\bibfield  {title} {\bibinfo {title} {Gel-assisted formation of giant unilamellar vesicles},\ }\href {https://doi.org/10.1016/j.bpj.2013.05.024} {\bibfield  {journal} {\bibinfo  {journal} {Biophysical Journal}\ }\textbf {\bibinfo {volume} {105}},\ \bibinfo {pages} {154} (\bibinfo {year} {2013})}\BibitemShut {NoStop}%
\end{thebibliography}%

\end{document}

\preprint{APS/123-QED}

\title{Supporting Information : Autonomous Engulfment of Active Colloids by Giant Lipid Vesicles}% Force line breaks with \\

\author{Florent Fessler}
 \author{Martin Wittman}
 \author{Juliane Simmchen}
\author{Antonio Stocco}%

\date{\today}% It is always \today, today,

\maketitle

\appendix

\section{Effect of the white light illumination on the particle velocity}

\begin{figure}[h!]
\includegraphics[width=0.7\linewidth]{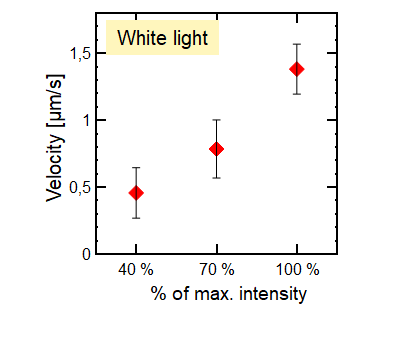}% Here is how to import EPS art
\caption{Average (on $N$= 10 trajectories) measured values for projected velocity $V$ for different illumination intensities (in percentage of maximum illumination intensity) at 100 mM glucose concentration. }
\label{fig1A}
\end{figure}

For a given concentration of glucose, changing the intensity of the white light illumination also modulates the measured projected active velocity $V$ from MSD fits. While the experiments conducted in this work were performed at low white light illumination intensity (40\% of maximum intensity), we show in Figure \ref{fig1A} that the average $V$ extracted for $N=10$ particles increase with increasing illumination intensity, as it was the case for blue light (see main text).

\section{Translational diffusion of isolated particles}

\subsubsection{Translational diffusion of free particles in glucose}

\begin{figure}[h!]
\includegraphics[width=0.7\linewidth]{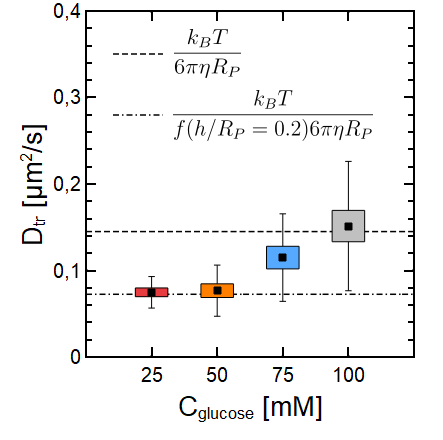}% Here is how to import EPS art
\caption{Average (over $N$= 10 trajectories for each point) of the values extracted from MSD fits for the projected velocity $V$ for different illumination intensities (in percentage of maximum illumination intensity) at 100 mM glucose concentration. Error bars account for the standard deviation.}
\label{fig2A}
\end{figure}

Using equation Eq.(1) from the main text, we can extract both the projected active velocity and the translational diffusion coefficient $D_{tr}$ of the Janus particles. Values for the extracted projected velocity $V$ are reported in the main text and $V$ increases with increasing glucose concentration. Here, Figure \ref{fig2A} shows the values for $D_{tr}$ as a function of the glucose concentration. We show that for lower glucose concentrations, the values are in agreement with Fàxen predictions of the diffusion coefficient of a particle close to a solid wall at a distance $h=0.2R_P=300 $ nm \cite{Goldmans1967}. For increasing glucose concentrations, $D_{tr}$ increases and becomes closer to the Stokes-Einstein prediction of the diffusion coefficient valid in the bulk. This suggests that the particle activity may have an impact on the swimming distance of the particle from the substrate or that additional short time noises due to the particle activity are present, which result into higher effective diffusion coefficients.

\subsubsection{Translational diffusion of fully wrapped particles}

\begin{figure}[t]
\includegraphics[width=0.75\linewidth]{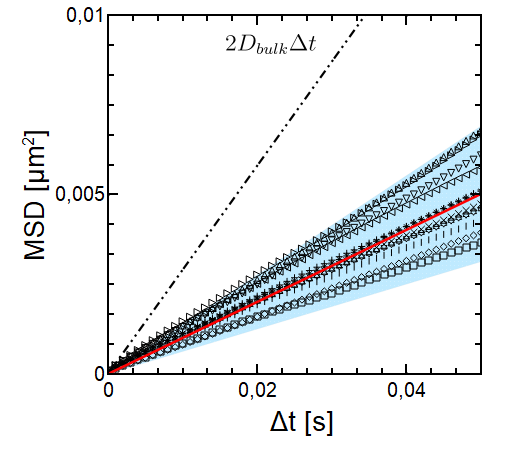}% Here is how to import EPS art
\caption{Experimental one-dimensional MSD plots for particles fully wrapped by the GUV (black symbols). The red curve shows the diffusion law $2\langle D_{wr}\rangle\Delta t$ with $\langle D_{wr}\rangle = 0.05 \ \mu$m$^2$.s$^{-1}$ the average value calculated from fits of the experimental data. The blue area covers $2\langle (D_{wr}\pm 1.5 \sigma_d)\rangle\Delta t$ where $\sigma_d$ is the standard deviation. Dashed black line shows the theoretical prediction for a spherical particle in the bulk according to the standard Stokes-Einstein relation.}
\label{fig3A}
\end{figure}

Once the particle has undergone the full wrapping transition, it remains stably wrapped by the lipid membrane and connected to the mother vesicle by a small neck. The evolution of the mean squared displacement with respect to lag time $\Delta t$ as shown in Figure \ref{fig3A} for wrapped particles indicates a severe slowing down of the translational diffusion as $\langle D_{wr} \rangle/D_{bulk} \approx 1/3$. An increase of the drag by a factor 3 can not be explained solely by the presence of a nearby wall. %Similarly, imputing this drag increase to the presence of a large water gap between the particle surface and the lipid membrane leading to a higher effective radius would lead to $R_{eff}=4.35 \mu$m which is not realistic. 
Given that the wrapped particle and the membrane neck form a connected object, additional dissipations arising from the membrane viscosity can contribute to the drag. The drag experienced by the neck is in fact equivalent to the drag of a disk embedded in a lipid membrane as modelled by Saffman-Delbr{\"u}ck \cite{Saffman1975}. Hence, the drag increase can be explained by this model, which also points to a very small gap between the particle and the wrapping membrane.

\section{Particle orientation tracking and correlation with motion persistence}

\begin{figure}[t]
\includegraphics[width=0.8\linewidth]{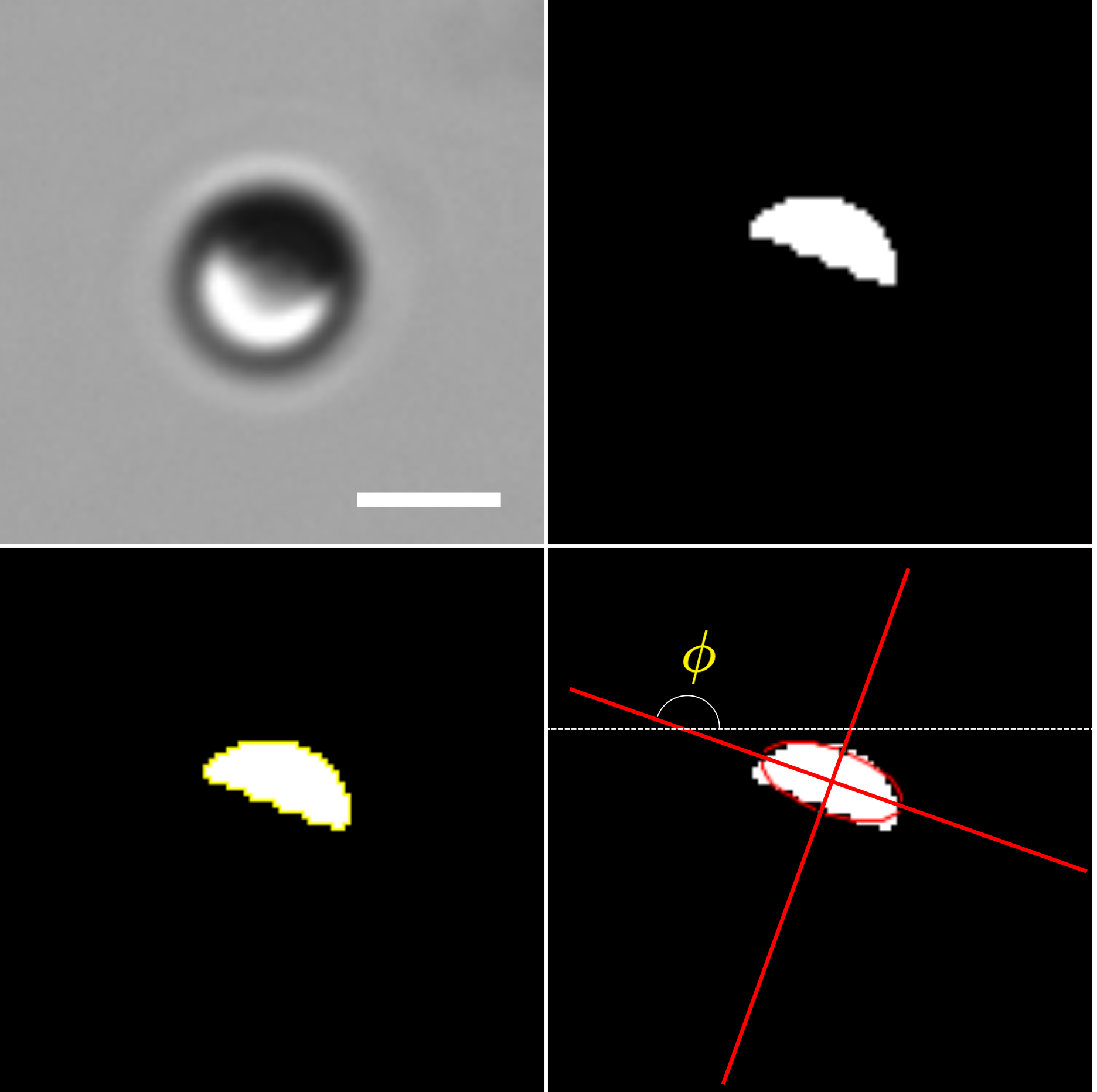}% Here is how to import EPS art
\caption{Procedure of orientation tracking using ImageJ. Top left panel shows the raw bright field image. Top right panel shows the same image but thresholded to obtain a binary image and subsequently inverted to have the dark cap appearing as maximum pixel value. Lower left panel shows the contour obtained using the "Wand" tool of ImageJ. Lower right panel shows the ellipse fitted (in red) to the contour, and the angle that makes the long axis of this ellipse with a fixed axis of the frame which we call $\phi$ (in-plane orientation).}
\label{figorient}
\end{figure}

Tracking of the orientation was performed using a method implemented previously for fluorescent Platinum coated particles. The method is depicted in Figure \ref{figorient} and will be exposed in the following. Using ImageJ, one can threshold the image in order to have a binary image with only the copper cap visible (which has lower pixel intensity than the background and the bare hemisphere of the Cu-SiO$_2$ Janus). Fitting an ellipse to the half-moon shaped cap allows to determine the angle between the long axis of the ellipse and a fixed axis on the image. An ImageJ routine allows to repeat this for each frame of a movie, providing an orientation trajectory. The associated tracking error is low provided that the particle out-of-plane orientation does not significantly deviates from $\alpha=\pi/2$, where $\alpha$ is the angle between the substrate normal and the normal  of the plane containing the Janus boundaries .

\begin{figure}[t]
\includegraphics[width=0.9\linewidth]{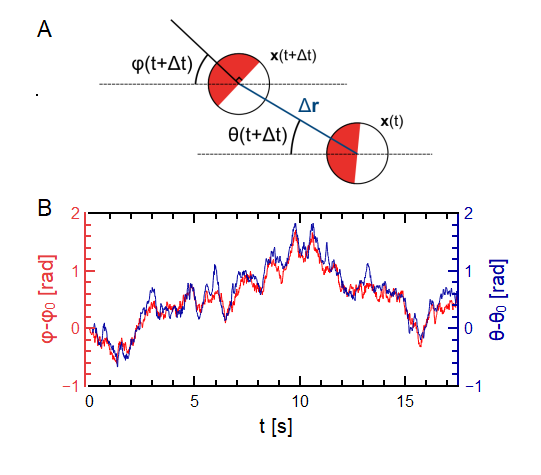}% Here is how to import EPS art
\caption{(a) Sketch defining the orientation $\phi$ and the direction of motion $\theta$ of a Janus swimmer. (b) Temporal evolution of $\phi-\phi_0$ and $\theta-\theta_0$ where $\phi_0=\theta_0=0$. The curve of $\theta-\theta_0$ was obtained by performing a sliding average over $N=20$ points to get rid of the noise arising from short times brownian motion and evidence the direction given by the active propulsion force.}
\label{figcorr}
\end{figure}

We can check both the robustness of this orientation tracking technique and the activity of Cu-SiO$_2$ particles in glucose by comparing the evolution of $\phi$ (the output angle of the tracking, see Figure \ref{figorient}) and $\theta$, which is the angle characterising the active motion persistence obtained from center-of-mass tracking of the particle (done using the tracking method exposed earlier) with a sliding average over 10 points to remove the Brownian translational noise. Figure \ref{figcorr} shows the good agreement between the observed direction of motion and the tracked orientation, confirming both the accuracy of the tracking method and the directed active propulsion of the particle in glucose.

\section{Rotational friction in the capture phase}

\begin{figure}[t]
\includegraphics[width=1\linewidth]{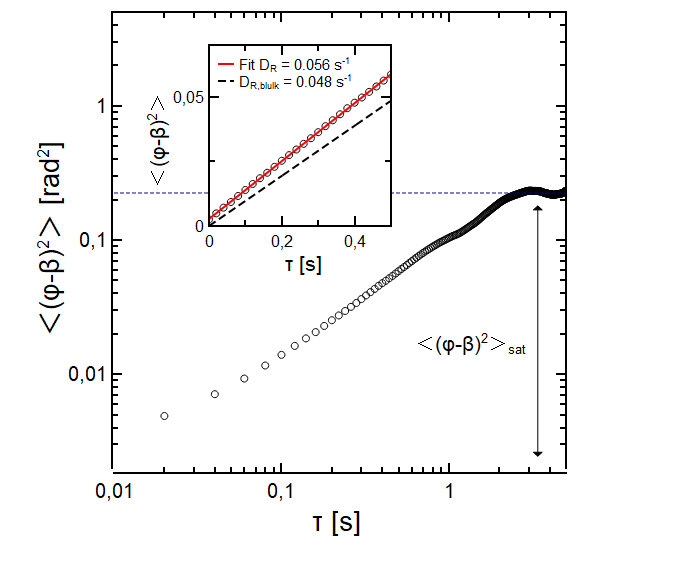}% Here is how to import EPS art
\caption{Log-log graph of the mean squared angular displacement during the capture phase, before the full wrapping transition. Short times (inset) allows to fit the diffusion coefficient $D_R$ while the plateau at large times evidences the angular confinement of the particle during this phase.}
\label{figdiff}
\end{figure}

During the capture phase, the angular confinement in the geometry of interest allows to track the orientation of the particle with the routine exposed previously. Figure \ref{figdiff} shows the mean squared angular displacement of a particle during the capture for a trajectory of 62 seconds with an acquisition frequency of 100 Hz yielding good statistics. The log-log scale graph allows to see the diffusive behavior at short times and confined orientational motion at long times with the appearance of a plateau of magnitude $\langle (\phi-\beta)^2\rangle = 0.22 $ rad$^2$. The linear scale inset shows how the measured linear diffusion (with $D_{tr}=0.056$ s$^{-1}$) compares to the theoretical rotational diffusion coefficient in the bulk.

\section{Comparison with pusher Pt-Si/H$_2$O$_2$ interaction with a GUV}

\begin{figure}[t]
\includegraphics[width=1\linewidth]{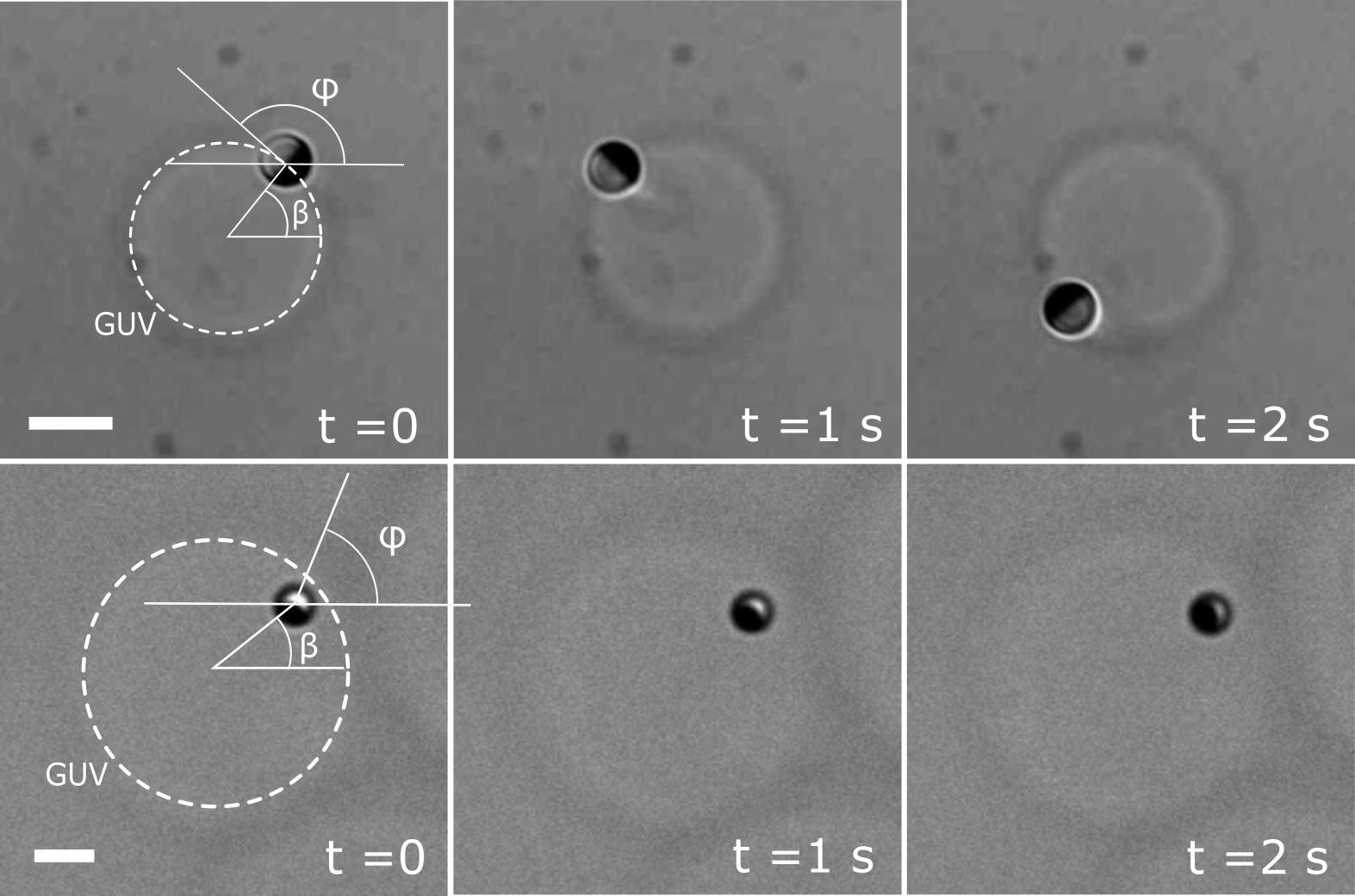}% Here is how to import EPS art
\caption{(Top row) Snapshots of a Pt-Si Janus puller swimmer in H$_2$O$_2$ at the periphery of a GUV for three times spaced of 1 second. (Bottom row) Snapshots of a Cu@SiO$_2$ Janus in glucose during the capture phase at the periphery of a GUV at three different times. }
\label{figcomp}
\end{figure}

The behavior of Cu-SiO$_2$ particles in glucose in the presence of giant unilamellar vesicles is very different from the one observed for the Pt-Si Janus colloids immersed in H$_2$O$_2$. This is shown in Figure \ref{figcomp} where the top row shows three snapshots taken at 1 s interval of a Pt-Si swimmer interacting with a GUV in H$_2$O$_2$. It clearly appears that the particle persistently orbits around the vesicle with $(\phi-\beta) \approx \pi/2$. The bottom row, instead, illustrates the case of a Cu-SiO$_2$ particle in glucose, which shows no persistent orbital motion and $(\phi-\beta) \approx 0$. The small variations of $\phi$ relatively to beta allow the particle to somewhat explore an arc of the GUV periphery. However, particle reorientation due to either hydrodynamic attraction with the GUV or Brownian motion results into an averaged zero net motion on the vesicle periphery.   

%\section{Contact curvatures for adhesion estimation}

%\begin{figure}[t]
%\includegraphics[width=1\linewidth]{SUPP_FIG/SUPP_3D.png}% Here is how to import EPS art
%\caption{(a) 3-dimensionnal reconstitution of confocal acquisitions for (top) POPC GUVs on a bare glass substrate and (bottom) a Copper covered glass substrate }
%\label{fig5}
%\end{figure}

\section{Stiffness and Diffusivity measurements}

\begin{figure}[b]
\includegraphics[width=1\linewidth]{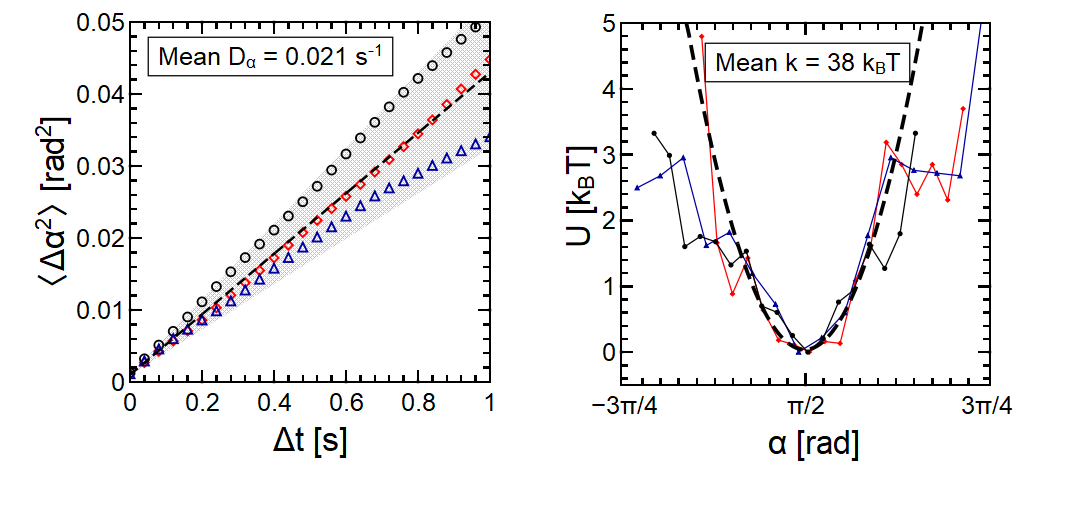}% Here is how to import EPS art
\caption{(a) Mean squared angular displacement curve obtained from the variations of $L$ for 3 trajectories with associated fits in plain lines. (b) Effective potential $U$ obtained from position histogram. Black dashed line shows the mean quadratic potential extracted from fits of the 3 potentials, giving the a stiffness value $k$. }
\label{figdiff2}
\end{figure}

To estimate the diffusivity $D_{\alpha}$ and stiffness $k$ that comes in the Kramers equation linking the hopping rate to the energy barrier $E_b$ (see main text), we use the data of the variations of $L$ during the capture phase. $L$ is acquired by tracking the position of both the particle center-of-mass and the vesicle center-of-mass using Blender. The variations of $L$ are translated in variations of the wrapping angle $\alpha$ ($\Delta L= R_P\Delta \alpha$). We can then compute the corresponding mean squared angular displacement of $\alpha$ and extract the diffusivity $D_{\alpha}$ from linear fits at short times since MSAD($\Delta t)=2D_{\alpha}\Delta t$, as shown in Figure \ref{figdiff2}. The average for three trajectories yields $\langle D_{\alpha} \rangle = 0.021 $ rad$^2$s$^{-1}$. 

The variations of $\alpha$ are confined in an effective potential, which we can obtain experimentally under the assumption that the probability to find a particle at a given position is linked to the energy by the standard Boltzmann weight. Assuming a quadratic potential, we can extract a stiffness $k$ by fitting the effective potential with a parabola. Average value $k=26.8 \ k_B T$ extracted from three parabolic fits is shown in Figure\ref{figdiff2} with the associated average parabola for the potential.

\section{Movies}

\label{videos}
\begin{itemize}
    \item S1 : MOVIE$\_$S1$\_$BRIGHTFIELD$\_$60x
    \item S2 : MOVIE$\_$S2$\_$FLUORESCENCE$\_$60x
\end{itemize}

\bibliography{apssamp}% Produces the bibliography via BibTeX.